\documentclass[twocolumn,english]{autart}    

\usepackage{graphicx}          
\usepackage{graphics}                               
\usepackage{color}
\usepackage[dvips]{epsfig}    
\usepackage{amsfonts}       
\usepackage{amssymb}
\usepackage{amsmath}
\usepackage{amssymb}
\usepackage{wrapfig}
\usepackage{bm}
\usepackage{subfig}
\usepackage{threeparttable}
\usepackage[wide]{sidecap}
\usepackage{array}
\usepackage{algorithm}
\usepackage{algpseudocode}
\usepackage{lipsum,babel}
\usepackage{color}
\usepackage{setspace}
\newtheorem{thrm}{Theorem}
\newtheorem{lemm}{Lemma}
\newtheorem{deff}{Definition}
\newtheorem{propo}{Proposition}
\newtheorem{assump}{Assumption}
\newtheorem{corol}{Corollary}
\newtheorem{conject}{Conjecture}
\newtheorem{remark}{Remark}

\newtheorem{proof}{Proof}

\def \vec {\mathrm{vec}}

\def \R {\mathbb{R}}
\def \C {\mathbb{C}}

\def \tailcoeff {\beta}
\def \tailexp {\alpha}

\def \policy {\pi}
\def \adaptivepolicy {\widehat{\pi}}
\def \optimalpolicy {\pi^\star}
\def \para {\theta}
\def \rrate {\gamma}
\def \sameorder {\asymp}
\def \randommatrix {\phi}
\def \normal {\boldsymbol{N}}
\def \Dnut {\mathbb{D}}

\newcommand{\Mnorm}[2]{{\left\vert\kern-0.35ex\left\vert\kern-0.35ex\left\vert #1 
		\right\vert\kern-0.35ex\right\vert\kern-0.35ex\right\vert}_{#2}}
\newcommand{\Opnorm}[3]{{\left\vert\kern-0.35ex\left\vert\kern-0.35ex\left\vert #1 
		\right\vert\kern-0.35ex\right\vert\kern-0.35ex\right\vert}_{#2 \to #3}}
\newcommand{\norm}[2]{{\left\vert\kern-0.35ex\left\vert #1 
		\right\vert\kern-0.35ex\right\vert}_{#2}}
\newcommand{\eigmax}[1]{\left| \lambda_{\max} \left( #1 \right)\right|}
\newcommand{\eigmin}[1]{\left| \lambda_{\min} \left( #1 \right)\right|}
\newcommand{\tr}[1]{\mathrm{tr} \left( #1 \right)}
\newcommand{\PP}[1]{\mathbb{P} \left(#1\right)}
\newcommand{\E}[1]{\mathbb{E} \left[#1\right]}

\newcommand{\rank}[1]{\mathrm{rank}\left(#1\right)}

\newcommand{\innerproductminconstant}[1]{\psi_0}

\newcommand{\regret}[2]{\mathcal{R}_{#1} \left(#2\right)}
\newcommand{\loss}[3]{\mathcal{L}_#1\left(#3\right)}

\newcommand{\dimension}[2]{\mathrm{dim}_{#1} \left(#2\right)}
\newcommand{\Lipschitz}[1]{\beta_{#1}}
\newcommand{\paraspace}[1]{\mathcal{X}_{#1}}
\newcommand{\tempparaspace}[1]{\Gamma_{#1}}
\newcommand{\Kmatrix}[1]{K\left(#1\right)}
\newcommand{\Lmatrix}[1]{L\left(#1\right)}
\newcommand{\extendedLmatrix}[1]{\widetilde{L}\left(#1\right)}

\newcommand{\avecost}[2]{\overline{\mathcal{J}}_{#2}\left(#1\right)}
\newcommand{\optcost}[1]{\mathcal{J}^\star \left(#1\right)}
\newcommand{\instantcost}[2]{c_{#1}\left(#2\right)}

\newcommand{\order}[1]{O \left(#1\right)}

\newcommand{\orderinv}[1]{\Omega \left(#1\right)}

\newcommand{\optpara}[1]{\widetilde{\para}_{#1}}

\newcommand{\estpara}[1]{\widehat{\para}_{#1}}
\newcommand{\estA}[1]{\widehat{A}_{#1}}

\newcommand{\trans}[1]{D_{#1}}

\newcommand{\levelset}[1]{\mathcal{S}\left(#1\right)}
\newcommand{\nullspace}[1]{\mathcal{N}\left(#1\right)}
\newcommand{\equilib}[1]{\mathcal{U}\left(#1\right)}
\newcommand{\optstate}[1]{x^{\star}\left(#1\right)}

\newif\ifarxiv
\arxivtrue

\begin{document}

\begin{frontmatter}
	
	\title{On Adaptive Linear-Quadratic Regulators}
	
	
	\author[]{Mohamad Kazem Shirani Faradonbeh, Ambuj Tewari, and George Michailidis}
	
\begin{abstract}
	Performance of adaptive control policies is assessed through the regret with respect to the optimal regulator, which reflects the increase in the operating cost due to uncertainty about the dynamics parameters. However, available results in the literature do not provide a quantitative characterization of the effect of the unknown parameters on the regret. Further, there are problems regarding the efficient implementation of some of the existing adaptive policies. Finally, results regarding the accuracy with which the system's parameters are identified are scarce and rather incomplete.
	
	This study aims to comprehensively address these three issues. First, by introducing a novel decomposition of adaptive policies, we establish a {\em sharp} expression for the regret of an {\em arbitrary} policy in terms of the deviations from the optimal regulator. Second, we show that adaptive policies based on slight modifications of the Certainty Equivalence scheme are efficient. Specifically, we establish a regret of (nearly) square-root rate for two families of randomized adaptive policies. The presented regret bounds are obtained by using {\em anti-concentration} results on the random matrices employed for randomizing the estimates of the unknown parameters. Moreover, we study the minimal additional information on dynamics matrices that using them the regret will become of logarithmic order. Finally, the rates at which the unknown parameters of the system are being identified are presented.
\end{abstract}

\end{frontmatter}



\section{Introduction}
This work studies the problem of designing {\em adaptive} policies for the following Linear-Quadratic (LQ) system. {Given an initial state $x(0) \in \R^{p}$, the system evolves as 
\begin{eqnarray}
x(t+1) &=& A_0x(t)+B_0u(t)+ w(t+1), \label{systemeq1} 
\end{eqnarray}
for $t \geq 0$, where the vector $x(t) \in \R^p$ corresponds to the state (and also output) of the system at time $t$, $u(t) \in \R^r$ is the control input, and $\left\{ w(t) \right\}_{t=1}^\infty$ denotes a sequence of random disturbances. Further, the instantaneous quadratic cost of the control law $\adaptivepolicy$ is denoted by 
\begin{eqnarray}
\instantcost{t}{\adaptivepolicy}&=& x(t)'Qx(t) + u(t)'Ru(t), \label{systemeq2}
\end{eqnarray}
where $Q \in \R^{p \times p}$, $R \in \R^{r \times r}$ are symmetric positive definite matrices, and $x(t)',u(t)'$ denote the transpose of the vectors $x(t),u(t)$.} The dynamics of the system, i.e., both the transition matrix $A_0 \in \mathbb{R}^{p \times p}$, as well as the input matrix $B_0 \in \mathbb{R}^{p \times r}$, are fixed and {\em unknown}, while $Q, R$ are assumed known. The overall objective is to adaptively regulate the system in order to minimize its long-term average cost. 

Although regulation of LQ systems represents a canonical problem in optimal control, adaptive policies have not been adequately studied in the literature. In fact, a large number of classical papers focuses on the setting of adaptive tracking,
where the objective is to steer the system to track a reference trajectory~\cite{lai1986extended,lai1986asymptotically,guo1988convergence,chen1989convergence,kumar1990convergence,lai1991parallel,guo1991aastrom,bercu1995weighted,guo1995convergence}. So, because the operating cost is not directly a function of the control signal (i.e., $R=0$), analysis of adaptive regulators becomes different and less technically involved. Therefore, existing results are not applicable to general LQ systems, wherein both the state and the control input impact the operating cost. The adaptive Linear-Quadratic Regulators (LQR) problem has been studied in the literature \cite{campi1998adaptive,duncan1999adaptive,bittanti2006adaptive,abbasi2011regret,ibrahimi2012efficient,faradonbeh2017finite,abeille2017thompson,ouyang2017control}, but there are still gaps that the present work aims to fill by addressing cost optimality, parameter estimation, and the trade-off between identification and control.

Since the system's dynamics are unknown, learning the key parameters $A_0,B_0$ is needed for designing an optimal regulation policy. However, the system operator needs to
apply some control inputs, in order to collect data (observations) for parameter estimation. A popular approach to design an adaptive regulator is Certainty Equivalence (CE)~\cite{bar1974dual}. Intuitively, its prescription is to apply a control policy {\em as if} the estimated parameters are the true ones guiding the system's evolution. In general, the inefficiency (as well as the inconsistency) of CE \cite{bittanti2006adaptive,lai1982least,becker1985adaptive} has led researchers to consider several modifications of the CE approach.

One idea is to use the principle of Optimism in the Face of Uncertainty (OFU)~\cite{abbasi2011regret,ibrahimi2012efficient,faradonbeh2017finite} (also known as bet on the best \cite{bittanti2006adaptive}, and the cost-biased approach~\cite{campi1998adaptive}). OFU recommends to apply the optimal regulators by treating {\em optimistic} approximations of the unknown matrices as the true dynamics~\cite{lai1985asymptotically}. Another idea is to replace the point estimate of the system parameters by a posterior distribution which is obtained through Bayes law by integrating a prior distribution and the likelihood of the data collected so far. One then draws a sample from this posterior distribution and applies the optimal policy, {\em as if} the system evolves according to the sampled dynamics matrices. This approach is known as Thompson (or posterior) sampling \cite{abeille2017thompson,ouyang2017control}. 

Note that most of the existing work in the literature is purely asymptotic in nature so that it establishes the convergence of the adaptive \emph{average} cost to the optimal value. It includes adaptive LQRs based on the OFU principle~\cite{campi1998adaptive,bittanti2006adaptive}, as well as those based on the method of random perturbations being applied to continuous time Ito processes~\cite{duncan1999adaptive}. However, results on the speed of convergences are rare and rather incomplete. On the other hand, from the identification viewpoint, consistency of parameter estimates is lacking for general dynamics matrices~\cite{polderman1986necessity,polderman1986note}. Moreover, accuracy rates for estimation of system parameters are only provided for minimum-variance problems~\cite{bercu1995weighted,guo1995convergence}. Indeed, the estimation rate for matrices describing the system's dynamics is not currently available for general LQ systems. 

Since in many applications the effective horizon is finite, the aforementioned asymptotic analyses are practically less relevant. Thus, addressing the optimality of an adaptive strategy under more sensitive criteria is needed. For this purpose, one needs to comprehensively examine the {\em regret}; i.e., the cumulative deviation from the optimal policy. Regret analyses are thus far limited to recent work addressing OFU adaptive policies \cite{abbasi2011regret,ibrahimi2012efficient,faradonbeh2017finite}, and results for TS obtained under restricted conditions \cite{abeille2017thompson,ouyang2017control}. One issue with OFU is the computational intractability of finding an optimistic approximation of the true parameters, since it needs to solve lots of non-convex matrix optimization problems. More importantly, we show that the existing regret bounds~\cite{abbasi2011regret,ibrahimi2012efficient,faradonbeh2017finite,abeille2017thompson,ouyang2017control} can be achieved or improved through simpler adaptive regulators. 

A key contribution of this work is a remarkably general result to address the performance of control policies. Namely, tailoring a novel method for regret decomposition, we utilize some results from martingale theory to establish Theorem~\ref{RegretTheorem}. It provides a {\em sharp} expression for the regret of arbitrary regulators in terms of the deviations from the optimal feedback. Leveraging Theorem~\ref{RegretTheorem}, we analyze two families of CE-based adaptive policies. 

First, we show that the growth rate of the regret is (nearly) square-root in time (of the interaction with the system), if the CE regulator is properly {\em randomized}. Performance analyses are presented for both common approaches of additive randomization and posterior sampling. Then, the adaptive LQR problem 
is discussed when additional information (regarding the unknown dynamics parameters of the system) is available. In this
case, a \emph{logarithmic} rate for the regret of generalizations of CE adaptive policies is established, assuming that the available side information satisfies an identifiability condition. Examples of side information include constraints on the rank or the support of dynamics matrices, that in turn lead to optimality of the linear feedback regulator, if the closed-loop matrix is accurately estimated. {Further, the identification performance of the corresponding adaptive regulators is also addressed. To the best of our knowledge, this work provides the first comprehensive study of CE-based adaptive LQRs, for both the identification and the regulation problem.} 

The remainder of the paper is organized as follows. The problem is formulated in Section~\ref{problem statement}. Then, we provide an expression for the regret of general adaptive policies in Subsection~\ref{exploitsubsection}. Subsequently, the consistency of estimating the dynamics parameter is given in Subsection~\ref{exploresubsection}. In Section~\ref{CE section}, we study the growth rate of the regret, as well as the accuracy of parameter estimation, for two randomization schemes. Finally, in Section \ref{CE with side} we study a general condition which leads to significant performance improvements in both regulation and identification. 
{\begin{remark}[Stochastic statements]
	All probabilistic equalities and inequalities throughout this paper hold almost surely, unless otherwise explicitly mentioned.
\end{remark}}
The following {notation} will be used throughout this paper. For a matrix $A \in \C^{k \times \ell}$, $A'$ denotes its transpose. When $k=\ell$, the smallest (respectively largest) eigenvalue of $A$ (in magnitude) is denoted by $\lambda_{\min} (A)$ (respectively $\lambda_{\max}(A)$). For $v \in \C^d$, define the norm $\norm{v}{} = \left( \sum\limits_{i=1}^{d} \left| v_i \right|^2 \right)^{1/2}$. We also use the following notation for the operator norm of matrices. For $A \in \C^{k \times \ell}$ let $\Mnorm{A}{} = \sup \limits_{\norm{v}{}=1} {\norm{Av}{}}$. In order to show the dimension of the manifold $\mathcal{M}$ we employ $\dimension{}{\mathcal{M}}$. 
Finally, to indicate the order of magnitude, we use $a_n = \order{b_n}$ whenever $\limsup\limits_{n \to \infty} \left| {a_n}/{b_n}\right| < \infty$\ifarxiv, employ $a_n = \orderinv{b_n}$ for $\liminf\limits_{n \to \infty} \left|{a_n}/{b_n}\right| >0$, and write $a_n \sameorder b_n$, as long as both $a_n = \order{b_n}, a_n = \orderinv{b_n}$ hold.\else.\fi 

\section{Problem Formulation} \label{problem statement}
We start by defining the adaptive LQR problem this work is addressing. The stochastic evolution of the system is governed by the dynamics \eqref{systemeq1}, where for all $t \geq 1$, $w(t)$ is the vector of random disturbances satisfying: 
\\$\E {w(t)}=0, \:\: \E {w(t)w(t)'}=C$, and $\eigmin{C}>0$.
\\For the sake of simplicity, the noise vectors $\left\{ w(t) \right\}_{t=1}^\infty$ are assumed to be independent over time $t$. The latter assumption is made to simplify the presentation, and generalization to martingale difference sequences (adapted to a filtration) is straightforward\footnote{It suffices to replace the involved terms with those consisting of the conditional expressions (w.r.t.\ the corresponding filtration).}. Further, the following moment condition for the noise process is assumed. 
\begin{assump}[Moment condition] \label{momentcondition}
	There is $\tailexp>4$, such that $\tailexp$-th moments exist: $\sup\limits_{t \geq 1} \E{\norm{w(t)}{}^{\tailexp}} < \infty$.
\end{assump}
In addition, we assume that the true dynamics of the underlying system are stabilizable, a minimal assumption for the optimal control problem to be well-posed. 
\begin{assump}[Stabilizability] \label{stabilizability}
The true dynamics $\left[A_0,B_0\right]$ is stabilizable:  there exists a stabilizing feedback $L \in \R ^{r \times p}$ such that $\eigmax{A_0+B_0L} < 1$.
\end{assump}
Note that Assumption \ref{stabilizability} implies stabilizability in the average sense: $\limsup\limits_{n \to \infty} n^{-1} \sum\limits_{t=0}^{n} \norm{x(t)}{}^2 < \infty$.
\begin{deff}
Henceforth, for $A \in \R^{p \times p},B \in \R^{p \times r}$, we use $\para$ to denote $\left[A,B\right]$. So, $\para \in \R^{p \times q}$, where $q=p+r$.
\end{deff} 

We assume {\em perfect} observations; i.e., the output of the system corresponds to the state vector $x(t)$. Next, an admissible control policy is a mapping $\policy$ that designs the input according to the dynamics matrices $A_0,B_0$, the cost matrices $Q,R$, and the history of the system:
\begin{eqnarray*}
	u(t) = \policy \left( A_0, B_0 , Q, R , \left\{ x(i) \right\}_{i=0}^t, \left\{ u(j) \right\}_{j=0}^{t-1} \right),
\end{eqnarray*}
for all $t\geq 0$. An {\em adaptive} policy such as $\adaptivepolicy$, is oblivious to the dynamics parameter $\para_0$; i.e.,
\begin{eqnarray*}
	u(t) = \adaptivepolicy \left( Q, R , \left\{ x(i) \right\}_{i=0}^t, \left\{ u(j) \right\}_{j=0}^{t-1} \right).
\end{eqnarray*}
When applying the policy $\policy$, the resulting instantaneous quadratic cost at time $t$ defined in \eqref{systemeq2} is denoted by $\instantcost{t}{\policy}$. For an arbitrary policy $\policy$, let $\avecost{A_0,B_0}{\policy}$ denote the expected average cost of the system: $\avecost{A_0,B_0}{\policy} = \limsup \limits_{n \to \infty} n^{-1} \sum \limits_{t=0}^{n-1} \E{\instantcost{t}{\policy}}$. Note that the dependence of $\avecost{\para_0}{\policy}$ to the known cost matrices $Q,R$ is suppressed. Then, the optimal expected average cost is defined as $\optcost{A_0,B_0} = \min \limits_{\policy} \avecost{A_0,B_0}{\policy}$,
where the minimum is taken over {\em all} admissible control policies. The following proposition provides an optimal policy for minimizing the average cost, based on the Riccati equations: 
	\begin{eqnarray} 
	\Kmatrix{\para} &=& Q + A'\Kmatrix{\para}A \notag \\
	&-& A' \Kmatrix{\para}B \left(B'\Kmatrix{\para}B+R\right)^{-1} B'\Kmatrix{\para}A , \label{ricatti2} \\
	\Lmatrix{\para} &=& -\left(B'\Kmatrix{\para}B+R\right)^{-1} B'\Kmatrix{\para}A. \label{ricatti1}
	\end{eqnarray}
Accordingly, define the linear time-invariant policy $\optimalpolicy$:
\begin{eqnarray} \label{optpolicydeff}
\optimalpolicy: \:\:\: u(t)= \Lmatrix{\para_0} x(t), \:\:\: t=0,1,2,\cdots.
\end{eqnarray}
\begin{propo}[Optimal policy \cite{chan1984convergence,de1986riccati,faradonbeh2018stabilization}] \label{stabilizable}
	If $\left[A_0,B_0\right]$ is stabilizable, \eqref{ricatti2} has a unique solution, and $\optimalpolicy$ defined in \eqref{optpolicydeff} is an optimal regulator. Conversely, if $\Kmatrix{\para_0}$ is a solution of \eqref{ricatti2}, $\Lmatrix{\para_0}$ defined by \eqref{ricatti1} is a stabilizer. 
\end{propo}
In the latter case of Proposition \ref{stabilizable}, the solution $\Kmatrix{\para_0}$ is unique and $\optimalpolicy$ is an optimal regulator. Note that although $\optimalpolicy$ is the only optimal policy among the time-invariant feedback regulators, there are uncountably many time varying optimal controllers. 

To rigorously set the stage, we denote the linear regulator $u(t)=L_t x(t)$ by $\policy= \left\{ L_t \right\}_{t=0}^\infty$, where $L_t$ is a $r \times p$ matrix determined according to $A_0, B_0, Q, R $, $\left\{ x(i) \right\}_{i=0}^t, \left\{ u(j) \right\}_{j=0}^{t-1}$. For time-invariant policy $\policy_0= \left\{ L_0 \right\}_{t=0}^\infty$, we use $\policy_0$ and $L_0$ interchangeably. For an adaptive operator, the dynamics matrices $A_0,B_0$ are unknown. Hence, adaptive policy $\adaptivepolicy= \left\{ \widehat{L}_t \right\}_{t=0}^\infty$ constitutes the linear feedbacks $u(t)=\widehat{L}_t x(t)$, where $\widehat{L}_t \in \R^{r \times p}$ is required to be determined according to $Q, R $, $ \left\{ x(i) \right\}_{i=0}^t, \left\{ u(j) \right\}_{j=0}^{t-1}$. In order to measure the efficiency of an arbitrary regulator $\policy$, the resulting instantaneous cost will be compared to that of the optimal policy $\optimalpolicy$ defined in \eqref{optpolicydeff}. Specifically, the {\em regret} of policy $\policy$ at time $n$ is defined as 
\begin{eqnarray} \label{regretdeff}
\regret{n}{\policy} = \sum \limits_{t=0}^{n-1} \left[ \instantcost{t}{\policy} - \instantcost{t}{\optimalpolicy} \right].
\end{eqnarray}
The comparison between adaptive control policies is made according to regret, which is the cumulative deviation of the instantaneous cost of the corresponding adaptive policy from that of the optimal controller $\optimalpolicy$. 

An analogous expression for regret is previously used for the problem of adaptive tracking \cite{lai1986extended,lai1986asymptotically}. An alternative definition of the regret that has been used in the existing literature \cite{abbasi2011regret,ibrahimi2012efficient,faradonbeh2017finite,abeille2017thompson,ouyang2017control} is the cumulative deviations from the optimal \emph{average} cost: $\sum\limits_{t=0}^{n-1} \left[ \instantcost{t}{\policy} - \optcost{\para_0} \right]$.
The expression above differs from $\regret{n}{\policy}$ by the term $\sum \limits_{t=0}^{n-1} \instantcost{t}{\optimalpolicy} - n \optcost{\para_0}$, which is studied in the following result.
{\begin{propo} \label{diffterm}
	We have
	\begin{eqnarray*}
		\limsup\limits_{n \to \infty} \frac{\sum\limits_{t=0}^{n-1} \instantcost{t}{\optimalpolicy} - n \optcost{\para_0}}{n^{1/2} \log n} < \infty.
	\end{eqnarray*}
\end{propo}}
Therefore, the aforementioned definitions for the regret are {\em indifferent}, as long as one can establish an upper bound of $\order{n^{1/2}}$ magnitude (modulo a logarithmic factor) for either definition. However, defining the regret by \eqref{regretdeff} leads to more accurate analyses and tighter results (e.g. the regret specification of Theorem~\ref{RegretTheorem}, and the logarithmic rate of Theorem~\ref{GCETheorem}). {To proceed, we introduce the following definition. 
\begin{deff}
For a stabilizable parameter $\para \in \R^{p \times q}$, define $\extendedLmatrix{\para}=\left[I_p , \Lmatrix{\para}'\right]'  \in \R^{q \times p}$. 
\end{deff}
We can then express the closed-loop matrices based on  $\para,\extendedLmatrix{\para}$. For arbitrary stabilizable $\para_1,\para_2$, if one applies the optimal feedback matrix $\Lmatrix{\para_1}$ to a system with dynamics parameter $\para_2$, the resulting closed-loop matrix is $A_2+B_2\Lmatrix{\para_1}=\para_2 \extendedLmatrix{\para_1}$.}

\section{General Adaptive Policies} \label{general theory}
Next, we study the properties of general adaptive regulators. First, we study the regulation viewpoint 
in Subsection \ref{exploitsubsection}, and examine the regret of arbitrary linear policies. Then, from an identification viewpoint, consistency of parameter estimation is considered in Subsection \ref{exploresubsection}.

\subsection{Regulation} \label{exploitsubsection}
The main result of this subsection provides an expression for the regret of an arbitrary (i.e., either adaptive or non-adaptive) policy. According to the following theorem, the regret of the regulator $\left\{ L_t \right\}_{t=0}^\infty$ is of
the same order as the summation of the squares of the deviations of the linear feedbacks $L_t$ from $\Lmatrix{\para_0}$. Note that it is stronger than the previously known result that expressed the regret as the summation of the deviations from 
$\Lmatrix{\para_0}$ (not squared) \cite{abbasi2011regret,ibrahimi2012efficient,faradonbeh2017finite,abeille2017thompson,ouyang2017control}. As will be shown shortly, this difference changes the nature of both the lower-bound, as well as the upper-bound of the regret. 
\begin{thrm}[Regret specification] \label{RegretTheorem}
	{Suppose that $\policy= \left\{ L_t \right\}_{t=0}^\infty$ is a linear policy. Letting $\left\{ \optstate{t} \right\}_{t=0}^\infty$ be the trajectory under the optimal policy $\optimalpolicy$, we have
	\begin{eqnarray*}
		0 < \liminf\limits_{n \to \infty} \frac{\regret{n}{\policy}}{\chi_n + \varrho_n} \leq \limsup\limits_{n \to \infty} \frac{\regret{n}{\policy}}{\chi_n + \varrho_n} < \infty,
	\end{eqnarray*}
	where $\varrho_n=\optstate{n}' \Kmatrix{\para_0} \optstate{n} - x(n)' \Kmatrix{\para_0} x(n)$, and $\chi_n=\sum\limits_{t=0}^{n-1} \norm{ \left(\Lmatrix{\para_0} - L_t \right) x(t) }{}^2$.}
\end{thrm}
{The above specification for the regret is remarkably general, since policy $\policy$ does not need to satisfy any condition. Even for destabilized systems, the exponential growth of the state (and so the regret) is captured by $\chi_n$. Conceptually, $\chi_n$ captures the effect of the \emph{past} sub-optimality $\left\{ L_t \right\}_{t=0}^{n-1}$ on the regret, while the influence of the sub-optimal feedback $\left\{ L_t \right\}_{t=n}^{\infty}$ to be applied henceforth is reflected in $\varrho_n$. This is formally stated in the following result, which also addresses the magnitude of $\norm{\optstate{n}}{}$. According to Assumption~\ref{momentcondition}, Corollary~\ref{RegretCorollary} shows that $\limsup\limits_{n \to \infty} n^{-1/2} \varrho_n =0 $.}
\begin{corol} \label{RegretCorollary}
	{We have $\limsup\limits_{n \to \infty} n^{-\tailcoeff} \norm{\optstate{n}}{}=0$, for all $\tailcoeff > 1/\tailexp$. Further, letting $L_t=\Lmatrix{\para_0}$ for $t\geq n$, and $\policy= \left\{ L_t \right\}_{t=0}^\infty$, we get 
	$0 < \regret{\infty}{\policy} / \chi_\infty < \infty$.}
\end{corol}
Theorem~\ref{RegretTheorem} can be used for the sharp specification of the performance of adaptive regulators. The immediate consequence of Theorem \ref{RegretTheorem} provides a tight upper bound for the regret of an adaptive policy, in terms of the linear feedbacks. Indeed, since the presented result is bidirectional and not just an upper bound, it will also provide a general information theoretic lower bound for the regret of an adaptive regulator. For stabilized dynamics, it is shown that the smallest estimation error when using a sample of size $t$ is at least of the order $t^{-1/2}$ \cite{simchowitz2018learning}. Thus, at time $t$, the error in the identification of the unknown dynamics parameter $\para_0$ is {\em at least} of the same order. 
Therefore, for the minimax growth rate of the regret, Theorem \ref{RegretTheorem} implies the lower bound $\log n$. 

In other words, for an arbitrary adaptive policy $\adaptivepolicy$, it holds that {$\liminf\limits_{n \to \infty} \left(\log n\right)^{-1} \regret{n}{\adaptivepolicy} > 0$}. In general, the information theoretic lower bound above is not known to be \emph{operationally} achievable because of the common trade-off between estimation and control. We will discuss the reasoning behind the presence of such a gap in Section \ref{CE section}, which leads to the operational lower bound {$\liminf\limits_{n \to \infty} n^{-1/2}\regret{n}{\adaptivepolicy} > 0$}. Nevertheless, in Section \ref{CE with side} we discuss settings where availability of some side information leads to an achievable regret of logarithmic order.

{Next, we provide some intuition behind Theorem~\ref{RegretTheorem} and Corollary~\ref{RegretCorollary}. The expression is in nature similar to the concept of memorylessness, as discussed below. The dynamics of the system in~\eqref{systemeq1} indicate that the influence of non-optimal control inputs lasts forever. That is, if $L_{t_1} x(t_1) \neq \Lmatrix{\para_0}x(t_1)$, then for all $t>t_1$, the state vector $x(t)$ deviates from the optimal trajectory $\left\{ \optstate{t} \right\}_{t=0}^\infty$, and future control inputs $\left\{u(t)\right\}_{t=t_1+1}^\infty$ can not fully {\em compensate} this deviation. However, according to Theorem~\ref{RegretTheorem}, the regret is dominated by the magnitude of the square of the deviations of the non-optimal feedbacks from $\Lmatrix{\para_0}$. In other words, if switching to the optimal feedback $\Lmatrix{\para_0}$ occurs, then the regret remains of the same order of the effect of the non-optimal control inputs previously applied, and so is memoryless.} 
 
\subsection{Identification} \label{exploresubsection}
Another consideration for an adaptive policy is the estimation (learning) problem. Since in general the operator has {\em no knowledge} regarding the dynamics parameter $\para_0$, a natural question to address is that of identifying $\para_0$, in addition to 
examining cost optimality. In this subsection, we address the asymptotic estimation consistency of general adaptive policies. That is, a rigorous formulation of the relationship between the {\em estimable} information (through observing the state of the system), and the \emph{desired} optimality manifold is provided.

On one hand, for a linear feedback $L$, the best one can do by observing the state vectors is ``closed-loop identification"~\cite{kumar1990convergence,faradonbeh2017finite}; i.e., estimating the closed-loop matrix $A_0+B_0L$ accurately. On the other hand, an adaptive policy is {\em at least} desired to provide a sub-linear regret;
\begin{eqnarray} \label{sublinearregreteq}
\limsup\limits_{n \to \infty} \frac{\regret{n}{\adaptivepolicy}}{n} = 0.
\end{eqnarray} 
The above two aspects of an adaptive policy provide the properties of the asymptotic uncertainty about the true dynamics parameter $\para_0$. By the uniqueness of $\Lmatrix{\para_0}$ according to Proposition \ref{stabilizable}, the linear feedbacks of the adaptive policy $\adaptivepolicy = \left\{ \widehat{L}_t \right\}_{t=0}^\infty$ require to converge to $\Lmatrix{\para_0}$. Further, $\adaptivepolicy$ uniquely identifies the asymptotic closed-loop matrix $\lim\limits_{t \to \infty} A_0+B_0 \widehat{L}_t$. This matrix according to \eqref{sublinearregreteq} is supposed to be $\para_0 \extendedLmatrix{\para_0}$. Putting the above together, the asymptotic uncertainty is reduced to the set of parameters $\para_\infty$ that satisfy
\begin{eqnarray} 
\Lmatrix{\para_\infty} = \Lmatrix{\para_0},\:\:\:\: \label{asympparaset1}
\para_\infty \extendedLmatrix{\para_0} = \para_0 \extendedLmatrix{\para_0}. \label{asympparaset2}
\end{eqnarray}
To rigorously analyze this uncertainty, we introduce some additional notation. First, for an arbitrary stabilizable 
$\para_{1}$, introduce the shifted null-space of the linear transformation $\extendedLmatrix{\para_1}: \R^{p \times q} \to \R^{p \times p}$ by $\nullspace{\para_{1}} $ as:
\begin{eqnarray} \label{nullspacedeff}
\nullspace{\para_{1}} = \left\{ \para \in \R^{p \times q} : \para \extendedLmatrix{\para_1} = \para_{1} \extendedLmatrix{\para_{1}} \right\}.
\end{eqnarray}
So, $\nullspace{\para_1}$ is indeed the set of parameters $\para$, such that the closed-loop transition matrix of two systems with dynamics parameters $\para,\para_1$ will be the same, if applying the optimal linear regulator in \eqref{ricatti1} calculated for $\para_1$. Hence, if the operator regulates the system by feedback $\Lmatrix{\para_1}$, one can not identify $\para,\para_1$. In other words, $\nullspace{\para_1}$ is the {\em learning capability} of adaptive regulators. Then, we define the {\em desired planning} of adaptive policies as follows. For an arbitrary stabilizable $\para_1$, define $\levelset{\para_{1}}$ as the level-set of the optimal controller function \eqref{ricatti1}, which maps $\para \in \R^{p \times q}$ to $\Lmatrix{\para} \in \R^{r \times p}$:
\begin{eqnarray} \label{levelsetdeff}
\levelset{\para_1} = \left\{ \para \in \R^{p \times q} : \Lmatrix{\para} = \Lmatrix{\para_1} \right\}.
\end{eqnarray}
Therefore, $\levelset{\para_1}$ is in fact the set of parameters $\para$, such that the calculation of optimal linear regulator \eqref{ricatti1} provides the same feedback matrix for both $\para,\para_1$. {Intuitively, $\nullspace{\para_0}$ reflects the identification aspect of the adaptive regulators by specifying the accuracy of the parameter estimation procedure. Similarly, $\levelset{\para_0}$ reflects the control aspect, and specifies the regulation performance in terms of optimality of the cost minimization procedure.} Hence, the asymptotic uncertainty about the true parameter $\para_0$ is according to \eqref{asympparaset2} limited to the set
\begin{eqnarray} \label{asymptoticuncertainty}
\mathcal{P}_0 = \levelset{\para_0} \cap \nullspace{\para_0}.
\end{eqnarray}
{The system theoretic interpretation is as follows. Assuming~\eqref{sublinearregreteq}, $\mathcal{P}_0$ is the smallest subset of dynamics parameters $\para$ that one can identify according to the state and the input sequences. Thus, the consistency of identifying the true dynamics parameter $\para_0$ is equivalent to $\mathcal{P}_0 =\left\{ \para_0 \right\}$. The following result establishes the properties of $\mathcal{P}_0$, and will be used later to discuss the operational optimality of adaptive regulators. It generalizes some results in the literature~\cite{polderman1986necessity,polderman1986note}.}
\begin{thrm}[Consistency] \label{adaptiveconsistency}
	The set $\mathcal{P}_0$ defined in \eqref{asymptoticuncertainty} is a shifted linear subspace of dimension $\dimension{}{\mathcal{P}_0} = \left( p - \rank{A_0} \right) r$.
\end{thrm}
Therefore, consistency of estimating $\para_0$ is {\em automatically} guaranteed for an adaptive policy with a sublinear regret, only if $A_0$ is a full-rank matrix. In other words, effective control (exploitation) suffices for consistent estimation (exploration) only if $\rank{A_0}=p$. For example, the sublinear regret bounds of OFU~\cite{abbasi2011regret,faradonbeh2017finite} imply consistency, assuming $A_0$ is of the full rank. Intuitively, a singular $A_0$ precludes unique identification of both of $A_0,B_0$ by~\eqref{asympparaset2}. Note that the converse is always true: consistency of parameter estimation implies the sublinearity of the regret. Clearly, full-rankness of $A_0$ holds for almost all $\para_0$ (with respect to Lebesgue measure).

\section{Randomized Adaptive Policies} \label{CE section}
The classical idea to design an adaptive policy is the following procedure known as CE. At every time $n$, its prescription is to apply the optimal regulator provided by \eqref{ricatti1}, as if the estimated parameter $\estpara{n}$ coincides exactly with the truth $\para_0$. According to \eqref{systemeq1}, a natural estimation procedure is to linearly regress $x(t+1)$ on the covariates $x(t),u(t)$, using all observations collected so far; $0 \leq t \leq n-1$. Formally, the CE policy is $\left\{ \Lmatrix{\estpara{n}} \right\}_{n=1}^\infty$, where $\estpara{n}$ is a solution of the least-squares estimator using the data observed until time $n$. That is, 
\begin{eqnarray*}
	\estpara{n} = \arg\min\limits_{\para \in \R^{p \times q}} \sum\limits_{t=0}^{n-1} \norm{x(t+1)- \para \extendedLmatrix{\estpara{t}} x(t)}{}^2.
\end{eqnarray*}

The issue with CE is that it is capable of {\em adapting} to a {non-optimal} regulation. Technically, CE possibly fails to \emph{falsify} an incorrect estimation of the true parameter~\cite{bittanti2006adaptive}. Suppose that at time $n$, the hypothetical estimate of the true parameter is $\estpara{n} \neq \para_0$. When applying the linear feedback $\Lmatrix{\estpara{n}}$, the true closed-loop transition matrix will be $\para_0 \extendedLmatrix{\estpara{n}}$. Then, if this matrix is the same as the (falsely) assumed closed-loop transition matrix $\estpara{n} \extendedLmatrix{\estpara{n}}$, the estimation procedure can fail to falsify $\estpara{n}$. So, if $\Lmatrix{\estpara{n}} \neq \Lmatrix{\para_0}$, the adaptive policy is not guaranteed to tend toward a better control feedback, and a non-optimal regulator will be persistently applied. 

Fortunately, if slightly modified, CE can avoid \emph{unfalsifiable} approximations of the true parameters. More precisely, we show that the set of unfalsifiable parameters defined below is of zero Lebesgue measure;
\begin{eqnarray} \label{unfalsedeff}
	\equilib{\para_0} = \left\{ \para \in \R^{p \times q}: \para_0 \extendedLmatrix{\para} = \para \extendedLmatrix{\para}  \right\}.
\end{eqnarray}
Note that by \eqref{nullspacedeff}, $\para_1 \in \equilib{\para_2}$ if and only if $\para_2 \in \nullspace{\para_1}$. Recalling the discussion in the previous section, $\nullspace{\para_1}$ captures the estimation ability of adaptive regulators. That is, the set $\equilib{\para_0}$ contains the matrices $\para$ for which the hypothetically assumed closed-loop matrix is indistinguishable from the true one. The next lemma sets the stage for the subsequent results which show that CE can be efficient, if it is suitably randomized.
\begin{lemm}[Unfalsifiable set] \label{unfalselemma}
	The set $\equilib{\para_0}$ defined in \eqref{unfalsedeff} has Lebesgue measure zero.
\end{lemm}
\subsection{Randomized Certainty Equivalence}
According to Lemma \ref{unfalselemma}, we can avoid the pathological set $\equilib{\para_0}$. As subsequently explained, it suffices to randomize the least-squares estimates of $\para_0$, with a small (diminishing) perturbation. First, such perturbations are chosen to be continuously distributed over the parameter space $\R^{p \times q}$, in order to evade $\equilib{\para_0}$. Further, since the linear transformation $\extendedLmatrix{\estpara{n}}$ is randomly perturbed, we can estimate the unknown dynamics parameter $\para_0$. Note that as discussed in the previous section, the sequence $\left\{\extendedLmatrix{\estpara{n}}\right\}_{n=0}^\infty$ relates the estimation of $\para_0$ to the accurate identification of the closed-loop matrix $\para_0 \extendedLmatrix{\estpara{n}}$. Finally, according to Theorem \ref{RegretTheorem}, the magnitude of the random perturbation needs to diminish sufficiently fast. Indeed, while a larger magnitude perturbation helps to the improvement of estimation, an efficient regulation requires it to be sufficiently small. Addressing this trade-off is the common dilemma of adaptive control. At the end of this section, we will examine this trade-off based on properties of estimation methods and the tight specification of the regret in Theorem \ref{RegretTheorem}.

In the sequel, we present the {\em Randomized Certainty Equivalence} (RCE) adaptive regulator. RCE is an episodic algorithm as follows. First, when identifying a linear dynamical system using $n$ observations, the estimation accuracy scales at rate $n^{-1/2}$. Therefore, one can defer updating of the parameter estimates until collecting sufficiently more data. This leads to the episodic adaptive policies, where the linear feedbacks are updated \emph{only} after episodes of exponentially growing lengths~\cite{faradonbeh2017finite}. In RCE, the randomization of the parameter estimate is episodic as well. Thus, calculation of the linear feedbacks $\Lmatrix{\estpara{n}}$ by \eqref{ricatti1} will occur sparsely (only $\order{\log n}$ times, instead of $n$ times), which remarkably reduces the computational cost of the algorithm.

\begin{algorithm}
	\caption{{\bf: RCE} } \label{RCEcode}
	\begin{algorithmic}
		\State Input: $\rrate >1$, and $\sigma_0>0$
		\State Let $\Lmatrix{\estpara{0}}$ be a stabilizer 
		\For{$m=0,1,2,\cdots$}
		\While{$n < \lfloor \rrate^m \rfloor$}
		\State Apply $u(n)=\Lmatrix{\estpara{n}} x(n)$ 
		\State $\estpara{n+1}=\estpara{n}$
		\EndWhile
		\State Update the estimate $\estpara{n}$ by \eqref{RCEalgo}
		\EndFor
	\end{algorithmic}
\end{algorithm}
To formally define RCE, let $\left\{\randommatrix_m\right\}_{m=0}^\infty$ be a sequence of i.i.d.\ $p \times q$ random matrices with independent $\normal \left(0,\sigma_0^2\right)$ entries, for a fixed $\sigma_0>0$. This sequence will be used to randomize the estimates. RCE has an arbitrary parameter $\rrate>1$ for determining the lengths of the episodes, and starts by an arbitrary initial estimate $\estpara{0}$ such that $\Lmatrix{\estpara{0}}$ stabilizes the system. To find such initial estimates, one can employ the existing adaptive algorithm to stabilize the system in a short period~\cite{faradonbeh2018stabilization}. Later on, we will briefly discuss the aforementioned stabilization algorithm. Then, for each time $n\geq 0$, we apply the linear feedback $\Lmatrix{\estpara{n}}$. If $n$ satisfies $n= \lfloor \rrate^m \rfloor$ for some $m \geq 0$, we update the estimate by  
\begin{eqnarray} \label{RCEalgo}
	\estpara{n} = \optpara{n}+ \arg\min\limits_{\para \in \R^{p \times q}} \sum\limits_{t=0}^{n-1} \norm{x(t+1)- \para \extendedLmatrix{\estpara{t}} x(t)}{}^2 ,\:\:\:\:\:\:\:\:\:
\end{eqnarray}
where $\optpara{n}= \left({n^{-1/4}}{\log^{1/4} n}\right) \randommatrix_m$ is the random perturbation. Otherwise, for $n \neq \lfloor \rrate^m \rfloor$, the policy does not update the estimates: $\estpara{n}=\estpara{n-1}$.
\begin{table*}
	{\small \begin{eqnarray}
		A_0 &=&\begin{bmatrix}
		1.04&         0&   -0.27\\
		0.52&   -0.81&    0.83\\
		0 &   0.04&   -0.90
		\end{bmatrix}, \:\:\:\:
		B_0 =\begin{bmatrix}
		-0.47&    0.61&   -0.29\\
		-0.50&    0.58&    0.25\\
		0.29&         0  & -0.72
		\end{bmatrix}, \:\:\:\:
		Q = \begin{bmatrix}
		0.65&   -0.08&   -0.14\\
		-0.08&    0.57&    0.26\\
		-0.14&    0.26&    2.50
		\end{bmatrix},\:\:\:\:
		R=\begin{bmatrix}
		0.20&    0.05&    0.08\\
		0.05&    0.14&    0.04\\
		0.08&    0.04&    0.24
		\end{bmatrix}. \label{simulationmatrices}
		\end{eqnarray}}
\end{table*}
Note that since the distribution of $\optpara{n}$ over $p \times q$ matrices is absolutely continuous with respect to Lebesgue measure, $\estpara{n}$ is stabilizable (as well as controllable \cite{bertsekas1995dynamic,faradonbeh2016optimality}). Therefore, by Proposition~\ref{stabilizable}, the adaptive feedback $\Lmatrix{\estpara{n}}$ is well defined.
\begin{remark} [\small Non-Gaussian Randomization] \label{perturbmomentcondition}
	In general, it suffices to draw $\left\{\randommatrix_m\right\}_{m=0}^\infty$ from an arbitrary distribution with bounded probability density functions on $\R^{p \times q}$ such that $\sup\limits_{m \geq 1} \E{\Mnorm{\randommatrix_m}{}^{4+\epsilon}} < \infty$, for some $\epsilon>0$.
\end{remark}
As mentioned before, the rate $\rrate$ determines the lengths of the episodes during which the algorithm uses $\estpara{n}$, before updating the estimate. Smaller values of $\rrate$ correspond to shorter episodes and thus more updates and additional randomization; i.e., the smaller $\rrate$ is, the better the estimation performance of RCE is. Although we will shortly see that such an improvement will not provide a better asymptotic rate for the regret, it speeds up the convergence and so is suitable if the actual time horizon is not very large. Further, it increases the number of times the Riccati equation \eqref{ricatti1} needs to be computed. Therefore, in practice the operator can decide $\rrate$ according to the time length of interacting with the system, and the desired computational complexity. It is important especially if the evolution of the real-world plant under control requires the feedback policy to be updated fast (compared to the time the operator needs to calculate the linear feedback). The following theorem addresses the behavior of RCE, and shows that adaptive policies based on OFU~\cite{abbasi2011regret,ibrahimi2012efficient,faradonbeh2017finite} do not provide a better rate for the regret, while they impose a large computational burden by requiring solving a matrix optimization problem. 
\begin{figure}[t!] 
	\centering
	\scalebox{.40}
	{\includegraphics{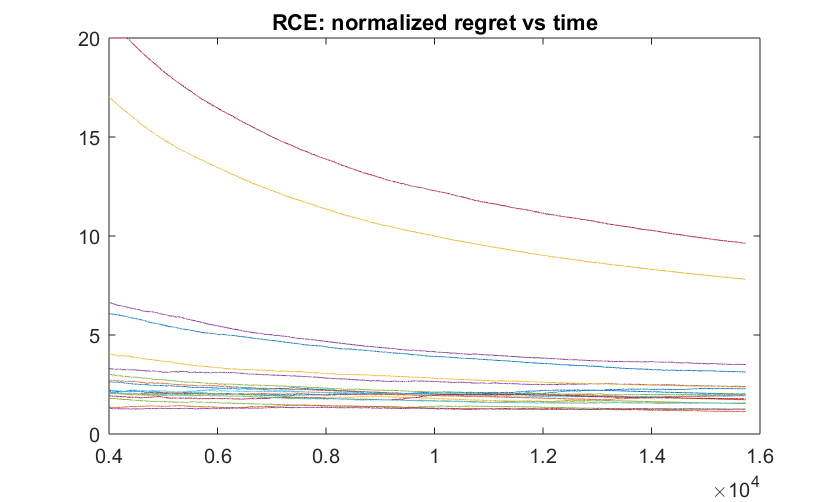}} 
	\scalebox{.40}
	{\includegraphics{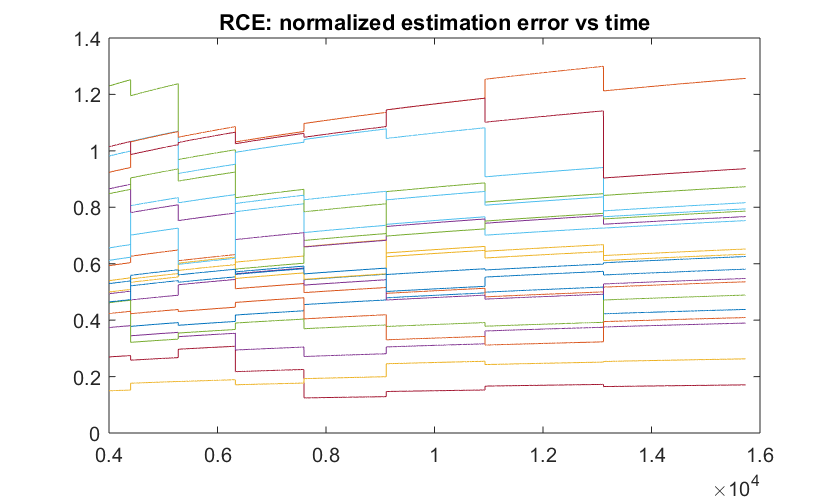}} 
	\caption{RCE performance: normalized regret $\left({n^{-1/2} \log^{-1} n} \right){\regret{n}{\adaptivepolicy}}$ vs $n$ (top), and normalized estimation error $\left({n^{1/4} \log^{-1/2} n }\right) {\Mnorm{\estpara{n}-\para_0}{}}$ vs $n$ (bottom).}
	\label{RCEFig}
\end{figure}
\begin{thrm}[RCE rates] \label{RCETheorem}
	Suppose that $\adaptivepolicy$ is RCE, and $\estpara{n}$ is the parameter estimate at time $n$. Then, we have
	\begin{equation*}
	\limsup\limits_{n \to \infty} \frac{\regret{n}{\adaptivepolicy}}{n^{1/2} \log n} < \infty, \limsup\limits_{n \to \infty} \frac{\Mnorm{\estpara{n}-\para_0}{}^2}{n^{-1/2} \log n}< \infty.
	\end{equation*}
\end{thrm}
Note that the analysis of RCE strongly leverages the specification of the regret presented in Theorem \ref{RegretTheorem}. Fig.~\ref{RCEFig} illustrates the results of Theorem~\ref{RCETheorem} by depicting the performance of RCE for $\rrate=1.2$, and the dynamics and cost matrices in~\eqref{simulationmatrices}. Curves of the normalized values of both the regret and the estimation error are depicted as a function of time, with the colors of the various curves corresponding to different replicates of the stochastic dynamics, as well as the adaptive policy RCE.
\subsection{Thompson Sampling}
Another approach in existing literature is {\em Thompson Sampling} (TS), which has the following Bayesian interpretation. Applying an initial stabilizing linear feedback, TS updates the estimate $\estpara{n}$ through posterior sampling. That is, the operator draws a realization $\estpara{n}$ of the Gaussian posterior for which the mean and the covariance matrix are determined by the data observed to date. 

Formally, let $\Sigma_0 \in \R^{q \times q}$ be a fixed positive definite (PD) matrix, and choose a coarse approximation $\mu_0 \in \R^{p \times q}$ of the truth $\para_0$. We will shortly explain an algorithmic procedure for computing such coarse approximations. Further, similar to RCE, fix the rate $\rrate>1$. Then, at each time $n \geq 0$, we apply $\Lmatrix{\estpara{n}}$, where $\estpara{n}$ is designed as follows. If $n$ satisfies $n= \lfloor \rrate^m \rfloor$ for some $m\geq 0$, $\estpara{n}$ is drawn from a Gaussian distribution $\normal \left( \mu_m,\Sigma_m^{-1}\right)$, where
\begin{eqnarray}
	\mu_m &=& \arg\min\limits_{\mu \in \R^{p \times q}} \sum\limits_{t=0}^{\lfloor \rrate^m \rfloor-1} \norm{x(t+1)- \mu \extendedLmatrix{\estpara{t}} x(t)}{}^2 , \label{TSalgo1}\\
	\Sigma_m &=& \Sigma_0 + \sum\limits_{t=0}^{\lfloor \rrate^m \rfloor-1} \extendedLmatrix{\estpara{t}} x(t) x(t)' \extendedLmatrix{\estpara{t}}' .  \label{TSalgo2}
\end{eqnarray}

\begin{figure}[t!] 
	\centering
	\scalebox{.4}
	{\includegraphics {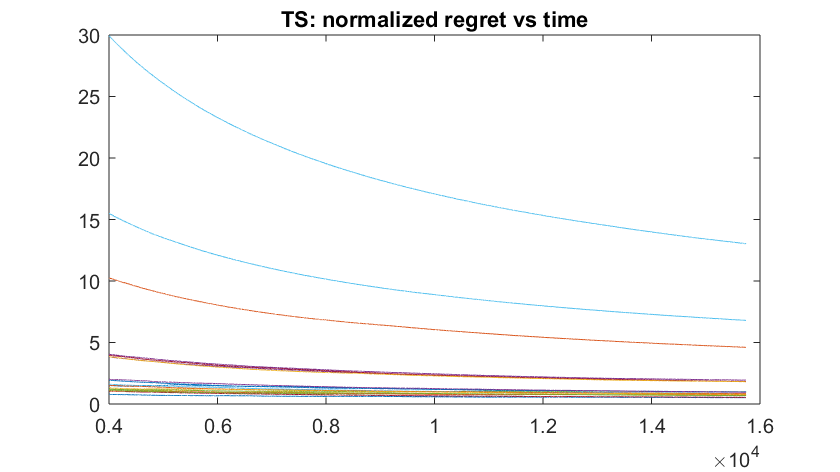}} 
	\scalebox{.4}
	{\includegraphics {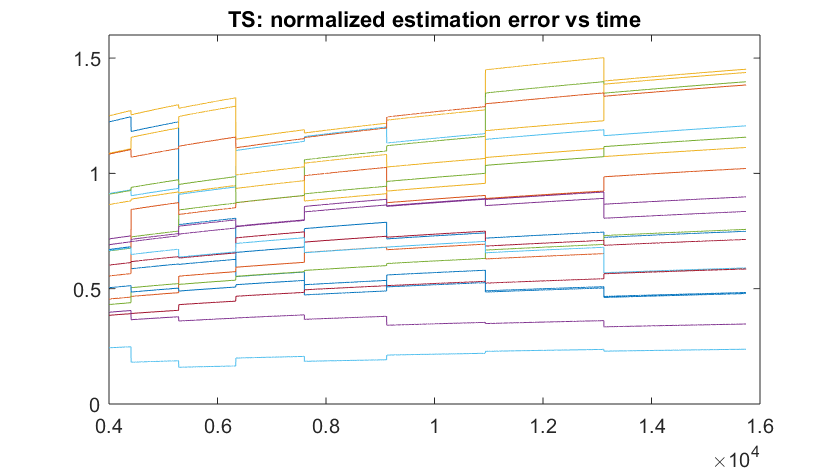}} 
	\caption{TS performance: normalized regret $\left({n^{-1/2} \log^{-1} n} \right){\regret{n}{\adaptivepolicy}}$ vs $n$ (top), and normalized estimation error $\left({n^{1/4} \log^{-1/2} n }\right) {\Mnorm{\estpara{n}-\para_0}{}}$ vs $n$ (bottom).}
	\label{TSFig}
\end{figure}

\begin{algorithm}
	\caption{{\bf : TS} } \label{TScode}
	\begin{algorithmic}
		\State Input: $\rrate >1$
		\State Let $\Sigma_0 \in \R^{q \times q}$ be PD, and $\Lmatrix{\estpara{0}}$ be a stabilizer 
		\For{$m=0,1,2,\cdots$}
		\While{$n < \lfloor \rrate^m \rfloor$}
		\State Apply $u(n)=\Lmatrix{\estpara{n}} x(n)$ 
		\State $\estpara{n+1}=\estpara{n}$
		\EndWhile
		\State Calculate $\mu_m,\Sigma_m$ by \eqref{TSalgo1}, \eqref{TSalgo2}
		\State Draw all rows of $\estpara{n}$ from $\normal \left( \mu_m,\Sigma_m^{-1}\right)$
		\EndFor
	\end{algorithmic}
\end{algorithm}
Namely, for $1 \leq i \leq p$, the $i$-th row of $\estpara{n}$ is drawn independently from a multivariate Gaussian distribution of mean $\mu_m^{(i)}$ (the $i$-th row of $\mu_m$), and covariance matrix $\Sigma_m^{-1}$. Otherwise, for $n \neq \lfloor \rrate^m \rfloor$ the policy does not update: $\estpara{n}=\estpara{n-1}$. Clearly, $\mu_m$ is the least-squares estimate and $\Sigma_m$ is the (unnormalized) empirical covariance of the data observed by the end of episode $m$. Note that unlike RCE, the randomization in TS is based on the state and control signals. The following result establishes the performance rates for TS.
\begin{thrm}[TS rates] \label{TSTheorem}
	Let the adaptive policy $\adaptivepolicy$ be TS, and the parameter estimate be $\estpara{n}$. Then, we have
	\begin{equation*}
	\limsup\limits_{n \to \infty} \frac{\regret{n}{\adaptivepolicy}}{n^{1/2} \log^2 n}< \infty ,
	\limsup\limits_{n \to \infty} \frac{\Mnorm{\estpara{n}-\para_0}{}^2}{n^{-1/2} \log^{2} n}< \infty.
	\end{equation*}
\end{thrm}
{Note that the above upper-bounds differ by those of Theorem~\ref{RCETheorem} by a logarithmic factor. The performance of TS for $\rrate=1.2$, and the matrices $A_0,B_0,Q,R$ in~\eqref{simulationmatrices} is depicted in Fig.~\ref{TSFig}. Clearly, the curves of the normalized regret and the normalized estimation error in Fig.~\ref{TSFig} fully reflect the rates of Theorem~\ref{TSTheorem}. For TS based adaptive LQRs, the {\em Bayesian regret} (i.e., the {\em expected} value of the regret, wherein the expectation is taken under the assumed prior) has been shown to be of a similar magnitude \cite{ouyang2017control}. Of course, this heavily relies on a Gaussian prior imposed on the true $\para_0$, and the (non-Bayesian) regret is known to be of $\order{n^{2/3}}$ magnitude~\cite{abeille2017thompson}. Therefore, Theorem \ref{TSTheorem} provides an improved regret bound for TS, thanks to Theorem \ref{RegretTheorem}. By assuming stronger assumptions (e.g. boundedness of the state), a similar result has been recently established for the case $p=1$, which holds uniformly over time~\cite{abeille2018improved}.}

{For the sake of completeness, we briefly discuss an existing adaptive stabilization procedure that one can employ before utilizing RCE or TS.  First, in the work of Faradonbeh et al.~\cite{faradonbeh2018stabilization}, it is shown that for some fixed $\epsilon_0>0$, a coarse approximation $\estpara{0}$ that satisfies $\Mnorm{\estpara{0}-\para_0}{} \leq \epsilon_0$, is sufficient for stabilizing the system~\cite{faradonbeh2018stabilization}. Note that the closed-loop matrix can be unstable before termination of an stabilization procedure. On the other hand, there exists a pathological subset of unstable matrices such that if the closed-loop transition matrix belongs to that subset, it is not feasible to be accurately estimated~\cite{faradonbeh2018finite}. Specifically, in order to ensure consistency, the true unstable closed-loop transition matrix during the stabilization period needs to be \emph{regular}, as defined below~\cite{faradonbeh2018finite}. The unstable square matrix $\trans{}$ is \emph{regular} if the eigenspaces corresponding to the eigenvalues of $\trans{}$ outside the unit circle are one dimensional~\cite{faradonbeh2018finite}. Then, it is established that random linear feedback matrices preclude the closed-loop irregularity~\cite{faradonbeh2018stabilization}. Therefore, the method of random feedback matrices guarantees that a coarse approximation of $\para_0$ is achievable in finite time, and a stabilization set can be constructed~\cite{faradonbeh2018stabilization}. Thus, we assume that the initial linear feedback matrix $\Lmatrix{\estpara{0}}$ is a stabilizer (i.e., $\eigmax{\para_0 \extendedLmatrix{\estpara{0}} } < 1$), and the system remains stable when RCE or TS is being employed. More details for establishing finite time adaptive stabilization are provided in the aforementioned reference~\cite{faradonbeh2018stabilization}. As a matter of fact, closed-loop regularity is not guaranteed, if only the control signals $\{ u(t) \}_{t=0}^\infty$ are randomized. Further, the classical framework of persistent excitation is not applicable due to the possible instability of the closed-loop matrix~\cite{faradonbeh2018finite,johnstone1983global,anderson1985adaptive,zhang1990further}.}

\subsection{Optimality}
Next, we discuss the reason for the presence of a significant gap between the operational regrets of Theorem~\ref{RCETheorem} and Theorem~\ref{TSTheorem}, and the information theoretic lower bound mentioned in Subsection \ref{exploitsubsection}. In fact, the following discussion shows that the logarithmic lower bound is {\em not} practically achievable. Nevertheless, in the next section
we show how using additional information for the true dynamics parameter yields a regret of logarithmic order. In the sequel, we discuss an argument that leads to the following conjecture: the regret is operationally of order $n^{1/2}$. For this purpose, we first state the following lemma about the level-set manifold $\levelset{\para_0}$ defined in \eqref{levelsetdeff}. It is a generalization of a previously established result for full-rank matrices \cite{polderman1986necessity,polderman1986note}.
 
\begin{lemm}[Optimality manifold] \label{optmanifold}
	The optimality level-set $\levelset{\para_0}$ is a manifold of dimension $\dimension{}{\levelset{\para_0}} = p^2 + \left( p -\rank{A_0} \right) \left( r - \rank{B_0} \right)$ at point $\para_0$.
\end{lemm}
By Theorem \ref{adaptiveconsistency}, we have $\dimension{}{\levelset{\para_0}}- \dimension{}{\mathcal{P}_0}=k$, where $k = p^2 - \left( p -\rank{A_0} \right) \rank{B_0}$. The tangent space of the manifold $\levelset{\para_0}$ at point $\para_0$, shares $\left( p -\rank{A_0}\right)r$ of its dimensions with $\nullspace{\para_0}$, and the other $k$ dimensions are apart from $\nullspace{\para_0}$. Intuitively, $\nullspace{\para_0}$ reflects the constraint of estimating the dynamics parameter, and $\levelset{\para_0}$ is the desired information to design an optimal policy. Thus, those $k$ dimensions of $\levelset{\para_0}$ which are not in $\nullspace{\para_0}$, {\em can not} be estimated unless the subspace $\nullspace{\para_0}$ is sufficiently perturbed. Such a perturbation is available only through applying non-optimal feedbacks, which yields a larger regret than the logarithmic rate mentioned in Subsection~\ref{exploitsubsection}. 

Next, we carefully analyze the regret based on the limits in falsifying the parameters not belonging to $\levelset{\para_0}$. First, inefficiency of an adaptive regulator compared to the optimal feedback $\Lmatrix{\para_0}$ is determined by the uncertainty for the exact specification of the optimality manifold $\levelset{\para_0}$. As an extreme example, suppose that $\levelset{\para_0}$ is provided to an operator who does not know $\para_0$. Then, denoting the adaptive policy above by $\adaptivepolicy$, we have $\regret{n}{\adaptivepolicy}=0$. Theorem \ref{RegretTheorem} states that if at time $n$ the adaptive regulator approximates $\levelset{\para_0}$ with error $\epsilon_n$, the growth in the regret is in magnitude $ \epsilon_n^2$. Thus, it suffices to examine the estimation accuracy $\epsilon_n$ that in turn depends both on the identification accuracy of the closed-loop transition matrix, as well as the falsification of dynamics parameters $\para \notin \levelset{\para_0}$.

{Now, suppose that the objective is to falsify $\para_1 \in \nullspace{\para_0}$, such that $\Mnorm{\para_1 - \para_0}{}= \sigma_n$, and $\para_1-\para_0$ is orthogonal to the linear manifold $\mathcal{P}_0$ defined in \eqref{asymptoticuncertainty}. The latter property of $\para_1$ dictates $\liminf\limits_{n \to \infty} \sigma_n^{-1} \Mnorm{\Lmatrix{\para_1}-\Lmatrix{\para_0}}{} > 0$. The key point is that in order to falsify $\para_1$, {\em non-optimal} linear feedbacks need to be applied {\em sufficiently many} times. For instance, if applying $\Lmatrix{\para_0}$, the estimation provides $\nullspace{\para_0}$, i.e., $\para_1$ can never get falsified. More generally, assume that $L$ is a $\delta_n$-perturbation of the optimal feedback: $\Mnorm{L- \Lmatrix{\para_0}}{}=\delta_n$. The shifted subspace of uncertainty when applying $L$ deviates from $\nullspace{\para_0}$ by at most $\order{\delta_n}$ (in the sense of inner products of the unit vectors). Next, assume that the operator applies $L$ (or a similar $\delta_n$-perturbed feedback) for a duration of $n$ time points. Note that the closed-loop estimation error is at least of the order of ${n^{-1/2}}$~\cite{simchowitz2018learning}. Thus, the operator can falsify $\para_1$ only if $\liminf\limits_{n \to \infty} n^{1/2} \delta_n \sigma_n > 0$. In other words, the adaptive regulator can avoid applying control feedbacks of distance at least ${n^{-1/2}\delta_n^{-1}}$ from the optimal feedback, only if control feedbacks of distance $\delta_n$ are in advance applied for a period of length $n$. Hence, we obtain $\liminf\limits_{n \to \infty} \sigma_n^{-2}\left(\regret{n+1}{\adaptivepolicy}-\regret{n}{\adaptivepolicy}\right)> 0$ by using Theorem \ref{RegretTheorem}, which also implies that such perturbed feedbacks impose a regret of the order $n \delta_n^2$. Putting together, we get $\liminf\limits_{n \to \infty} \regret{n}{\adaptivepolicy} \left(\regret{n+1}{\adaptivepolicy} - \regret{n}{\adaptivepolicy}\right)>0$. It leads to the following conjecture which constitutes an interesting direction for future work.}
\begin{conject}[lower bound]
	For an arbitrary adaptive policy $\adaptivepolicy$ we have 
	$\liminf\limits_{n \to \infty} n^{-1/2}\regret{n}{\adaptivepolicy}>0$.
\end{conject}
Note that if the above conjecture is true, RCE and TS provide a nearly optimal bound for the regret. Even the logarithmic gap between the lower and upper bounds is inevitable, due to the existence of an analogous gap in the closed-loop identification of linear systems \cite{simchowitz2018learning}. Further, the above discussion explains the intuition behind the design of RCE. Specifically, the magnitude of the perturbation $\Mnorm{\optpara{n}}{}$ according to the above discussion is optimally selected, since it satisfies $0 < \liminf\limits_{n \to \infty} \sigma_n^{-1} \delta_n \leq \limsup\limits_{n \to \infty} \sigma_n^{-1} \delta_n < \infty$, modulo a logarithmic factor. Indeed, if randomization is (significantly) smaller in magnitude than $n^{-1/4}$, the portion of the regret due to such a perturbation will reduce. However, it also reduces the accuracy of the parameter estimate. Thus, the other portion of the regret due to estimation error will increase. A similar discussion holds for larger magnitudes of the perturbation $\optpara{n}$. On the other hand, the magnitude of randomization in TS is determined by the collected observations. As one can see in the proof of Theorem \ref{TSTheorem}, a similar magnitude of randomization is automatically imposed by the structure of TS adaptive LQR.  

\section{Generalized Certainty Equivalence} \label{CE with side}
It is possible that the operator has additional information on the dynamics. Examples of such information are the set of non-zero entries of $\para_0$, the rank of $\para_0$, or a plant whose subsystems evolve independently of each other. Another example comes from large network systems, where a substantial portion of the matrix $\para_0$ entries
are zero~\cite{faradonbeh2016optimality}. Further, it is easy to see that the transition matrix of a system whose dynamics exhibit longer memory has a specific form~\cite{guo1991aastrom,faradonbeh2018finite}. 

In such cases, this additional structural information on $\para_0$ can be used by the operator in order to obtain a smaller regret for the adaptive regulation of the system. Nevertheless, a comprehensive theory needs to formalize how this side information can provide theoretical sharp bounds for the regret. In this section, we provide an identifiability condition that ensures that the adaptive LQRs attain the informational lower bound of logarithmic order. In addition to the classical CE adaptive regulator, we also consider the family of CE-based schemes which provide a logarithmic order of magnitude for the regret.

First, we introduce the {\em Generalized Certainty Equivalence} (GCE) adaptive regulator. GCE is an episodic algorithm with exponentially growing duration of episodes. Instead of randomizing the parameter estimate similar to RCE and TS, in GCE the least-squares estimate is perturbed with an arbitrary matrix $\optpara{n}$. Suppose that the operator knows that $\para_0 \in \tempparaspace{0}$, based on side information $\tempparaspace{0} \subset \R^{p \times q}$. Then, fixing the rate $\rrate>1$, at time $n\geq 0$, we apply the controller $\Lmatrix{\estpara{n}}$. If $n$ satisfies $n= \lfloor \rrate^m \rfloor$ for some $m\geq 0$, we update the estimate by
\begin{eqnarray} \label{GCEalgo}
	\estpara{n} &=& \optpara{n}+ \arg\min\limits_{\para \in \tempparaspace{0}} \sum\limits_{t=0}^{n-1} \norm{x(t+1)- \para \extendedLmatrix{\estpara{t}} x(t)}{}^2 ,\:\:
\end{eqnarray}
where $\optpara{n}$ is {\em arbitrary}, and satisfies $\limsup\limits_{n \to \infty} n^{1/2}\Mnorm{\optpara{n}}{}<\infty$. For $n \neq \lfloor \rrate^m \rfloor$ the policy does not update: $\estpara{n}=\estpara{n-1}$. Note that if $\optpara{n}=0$, we get the episodic CE adaptive regulator. To proceed, we define the following condition. 
\begin{algorithm}
	\caption{{\bf: GCE} } \label{GCEcode}
	\begin{algorithmic}
		\State Inputs: $\rrate >1$, $\tempparaspace{0} \subset \R^{p \times q}$
		\State Let $\Lmatrix{\estpara{0}}$ be a stabilizer 
		\For{$m=0,1,2,\cdots$}
		\While{$n < \lfloor \rrate^m \rfloor$}
		\State Apply $u(n)=\Lmatrix{\estpara{n}} x(n)$ 
		\State $\estpara{n+1}=\estpara{n}$
		\EndWhile
		\State Update the estimate $\estpara{n}$ by \eqref{GCEalgo}
		\EndFor
	\end{algorithmic}
\end{algorithm}
\begin{deff}[Identifiability]\label{Sidentifiability}
	Suppose that there is $\tempparaspace{0} \subset \R^{p \times q}$ such that $\para_0 \in \tempparaspace{0}$. Then, $\para_0$ is identifiable, if for some $\Lipschitz{0}< \infty$ and all stabilizable $\para_1,\para_2 \in \tempparaspace{0}$:
	\begin{eqnarray} \label{Sideq}
	\Mnorm{\Lmatrix{\para_2} - \Lmatrix{\para_0}}{} \leq \Lipschitz{0} \Mnorm{\left(\para_2 - \para_0\right) \extendedLmatrix{\para_1}}{}.
	\end{eqnarray}
\end{deff}
Intuitively, the definition above describes settings where side information $\tempparaspace{0}$ is sufficient in the sense that an $\epsilon$-accurate identification of the closed-loop matrix (the RHS of \eqref{Sideq}) provides an $\order{\epsilon}$-accurate approximation of the optimal linear feedback (the LHS of \eqref{Sideq}). Subsequently, we provide concrete examples of $\tempparaspace{0}$, such as presence of sparsity or low-rankness in $\para_0$. Essentially, a finite union of manifolds of proper dimension in the space $\R^{p \times q}$ suffices for identifiability. To see that, we use the critical subsets $\nullspace{\para_0}, \levelset{\para_0}$, and $\mathcal{P}_0$ defined in \eqref{nullspacedeff}, \eqref{levelsetdeff}, and \eqref{asymptoticuncertainty}, respectively. 

First, note that $\mathcal{P}_0 \subset \levelset{\para_0}$ provides the optimal linear feedback $\Lmatrix{\para_0}$. Hence, for $\para_1 \in \nullspace{\para_0}$, $\Mnorm{\Lmatrix{\para_1} - \Lmatrix{\para_0}}{}$ and $\inf\limits_{\para \in \mathcal{P}_0} \Mnorm{\para_1 - \para }{}$ are of the same order of magnitude. Then, according to Theorem~\ref{adaptiveconsistency}, both $\nullspace{\para_0}$ and $\mathcal{P}_0$ are shifted linear subspaces passing through $\para_0$. Since $\dimension{}{\nullspace{\para_0}}=pr$, the null-space $\nullspace{\para_0}$ shares $ \left( p - \rank{A_0} \right) r$ dimensions with $\mathcal{P}_0$, and has $\dimension{}{\nullspace{\para_0}} - \dimension{}{\mathcal{P}_0} = \rank{A_0} r$
dimensions orthogonal to $\mathcal{P}_0$. {The regret of an adaptive regulator $\adaptivepolicy$ becomes larger than a logarithmic function of time, because of the uncertainty ${\nullspace{\para_0}}/{\mathcal{P}_0}$. In other words, although the RHS of \eqref{Sideq} is estimated accurately, the aforementioned uncertainty precludes obtaining an accurate approximation for the LHS of \eqref{Sideq}. In Definition~\ref{Sidentifiability}, additional knowledge about $\para_0$ removes such uncertainty.} Thus, a manifold (or a finite union of manifolds) of dimension $pq - \rank{A_0} r$ implies the aforementioned identifiability condition. Below, we provide some examples of $\tempparaspace{0}$.

	(i) Optimality manifold: obviously, a trivial example is $\tempparaspace{0}=\levelset{\para_0}$. In this case, the LHS of \eqref{Sideq} vanishes.\\
	(ii) Support condition: let $\tempparaspace{0}$ be the set of $p \times q$ matrices with a priori known support $\mathcal{I}$. That is, for some set of indices $\mathcal{I} \subset \left\{ (i,j): 1 \leq i \leq p, 1 \leq j \leq q \right\}$, entries of all matrices $\para \in \tempparaspace{0}$ are zero outside of $\mathcal{I}$; $\tempparaspace{0}=\left\{ \para= \left[ \para_{ij} \right]: \para_{ij}=0 \text{ for } (i,j) \notin \mathcal{I} \right\}$. Then, $\tempparaspace{0}$ is a (basic) subspace of $\R^{p \times q}$ and can satisfy the identifiability condition \eqref{Sideq}. Note that it is necessary to have $\dimension{}{\tempparaspace{0}} = \left| \mathcal{I} \right| \leq pq - \rank{A_0}r$. \\
	(iii) Sparsity condition: let $\tempparaspace{0}$ be the set of all $p \times q$ matrices with at most $pq - \rank{A_0}r$ non-zero entries. Then, $\tempparaspace{0}$ is the union of the matrices with support $\mathcal{I}$ for different sets $\mathcal{I}$. Hence, the previous case implies that $\tempparaspace{0}$ is a finite union of manifolds of proper dimension.\\
	(iv) Rank condition: let $\tempparaspace{0}$ be the set of $p \times q$ matrices $\para$ such that $\rank{\para} \leq d$. Then, $\tempparaspace{0}$ is a finite union of manifolds of dimension at most $d \left( p+q-d \right)$ \cite{shalit2012online}. Hence, if $d \left( p+q-d \right) \leq pq - \rank{A_0} r$, and \eqref{Sideq} holds, $\para_0$ is identifiable.\\
	(v) Subspace condition: for $k= \rank{A_0}r$, let $\left\{\para_i\right\}_{i=1}^k$ be $p \times q$ matrices such that $\para_i \extendedLmatrix{\para_0}=0$. Suppose that $\para_1, \cdots, \para_k$ are linearly independent: if $\sum\limits_{i=1}^k a_i \para_i =0$, then $a_1=\cdots=a_k=0$. Define $\tempparaspace{0} = \left\{ \para+\para_0 : \tr{\para'\para_i}=0 \text{ for all } 1 \leq i \leq k \right\}$. If for all $1 \leq i \leq k$ it holds that $\para_0+\para_i \notin \mathcal{P}_0$, then $\tempparaspace{0}$ satisfies the identifiability condition of Definition \ref{Sidentifiability}.
	
The following Theorem establishes the optimality of GCE under the identifiability assumption. As mentioned in Section \ref{CE section}, a logarithmic gap between the lower and upper bounds for the regret is inevitable due to similar limitations in system identification \cite{simchowitz2018learning}.
\begin{thrm}[GCE Rates] \label{GCETheorem}
	Suppose that $\para_0$ is identifiable and the adaptive policy $\adaptivepolicy$ corresponds to GCE. Defining $\mathcal{P}_0$ by~\eqref{asymptoticuncertainty}, let $\estpara{n}$ be the parameter estimate at time $n$. Then, we have
	\begin{eqnarray*}
	\limsup\limits_{n \to \infty} \frac{ \regret{n}{\adaptivepolicy}}{\log^2 n} < \infty,
	\limsup\limits_{n \to \infty} \frac{\inf\limits_{\para \in \mathcal{P}_0} \Mnorm{\estpara{n}-\para}{}^2}{n^{-1} \log n} < \infty.
	\end{eqnarray*}
\end{thrm}
Comparing the above result with Theorem \ref{RCETheorem} and Theorem \ref{TSTheorem}, the identifiability assumption leads to significant improvements in rates of both the regret and the estimation error. Moreover, if $\rank{A_0}=p$, then $\mathcal{P}_0= \left\{ \para_0 \right\}$. Thus, the estimation accuracy in Theorem \ref{GCETheorem} becomes: $\limsup\limits_{n \to \infty} n \left(\log^{-1} n\right) \Mnorm{\estpara{n}- \para_0}{}^2< \infty$. Finally, Theorem~\ref{GCETheorem} improves an existing result for identifiable systems. That is, under stronger assumptions, Ibrahimi et al.~\cite{ibrahimi2012efficient} show the regret bound $\order{n^{1/2} \log^2 n}$ for adaptive policies based on OFU. However, according to Theorem~\ref{GCETheorem}, the regret of GCE is $\order{\log^2 n}$.

\section{Concluding Remarks}
The performances of adaptive policies for LQ systems is addressed in this work, including both aspects of regulation and identification. First, we established a general result which specifies the regret of an arbitrary adaptive regulator in terms of the deviations from the optimal feedback. This {\em tight} bidirectional result provides a powerful tool to analyze the subsequently presented policies. That is, we show that slight modifications of CE provide a regret of (nearly) square-root magnitude. The modifications consist of two basic approaches of randomization: additive randomness, and Thompson sampling. In addition, we formulated a condition which leads to {\em logarithmic} regret. The rates of {\em identification} are also discussed for the corresponding adaptive regulators. 

Rigorous establishment of the proposed operational {\em lower bound} for the regret is an interesting direction for future works. Besides, extending the developed framework to other settings such as {\em switching} systems, or those with {\em imperfect} observations are topics of interest. On the other hand, extensions to the dynamical models illustrating {\em network} systems (e.g., high-dimensional sparse dynamics matrices) is a challenging problem for further investigation.


\newpage
\appendix
\section{Proofs of Main Results}
{The proofs of the main theorems are given next. \ifarxiv Proofs \else Due to space limitations, proofs \fi of auxiliary lemmas are deferred to the \ifarxiv appendix.\else supplementary material~\cite{faradonbeh2018optimality}.\fi}
\subsection{\bf Proof of Theorem \ref{RegretTheorem} and Corollary~\ref{RegretCorollary}}
Given $n \geq 1$, and the linear policy $\policy= \left\{ L_t \right\}_{t=0}^{n-1}$, define the sequence of policies $\policy_0, \cdots, \policy_n$ as follows.
\begin{eqnarray*}
	\policy_0 &=& \left\{ \Lmatrix{\para_0}, \cdots, \Lmatrix{\para_0} \right\}, \\
	\policy_1 &=& \left\{ L_0, \Lmatrix{\para_0}, \cdots, \Lmatrix{\para_0} \right\}, \\
	\vdots \\
	\policy_n &=& \left\{ L_0, L_1, \cdots, L_{n-1} \right\}.
\end{eqnarray*}
Indeed, the policy $\policy_i$ applies the same feedback as $\policy$ at every time $t<i$, and then for $t \geq i$ switches to the optimal policy $\optimalpolicy$. Clearly, $\policy_0=\optimalpolicy$, and $\policy_n=\policy$. Since
\begin{eqnarray} \label{regretdecompos1}
\regret{n}{\policy}= \sum\limits_{k=1}^{n} \sum\limits_{t=0}^{n-1} \left[\instantcost{t}{\policy_k}-\instantcost{t}{\policy_{k-1}}\right], 
\end{eqnarray}
it suffices to find $\instantcost{t}{\policy_k}-\instantcost{t}{\policy_{k-1}}$, for $1 \leq k \leq n$, and $0 \leq t \leq n-1$. Fixing $k$, let $\left\{ x(t) \right\}_{t=0}^{n-1}, \left\{ y(t) \right\}_{t=0}^{n-1}$ be the state trajectories under $\policy_k, \policy_{k-1}$, respectively. So, letting $\Dnut=A_0+B_0 \Lmatrix{\para_0}$ and $D_{k-1}=A_0+B_0 L_{k-1}$, we have $x(t) = y(t)$ for $0 \leq t \leq k-1$, as well as $\instantcost{t}{\policy_k} = \instantcost{t}{\policy_{k-1}}$ for $0 \leq t \leq k-2$, and $x(k) = D_{k-1} x(k-1) + w(k)$. Further, if $k \leq t \leq n-1$, then
\begin{eqnarray*}
	y(t) &=& \Dnut^{t-k+1} x(k-1) + \sum\limits_{j=k}^t \Dnut^{t-j} w(j) ,\\
	x(t) &=& \Dnut^{t-k} D_{k-1} x(k-1) + \sum\limits_{j=k}^t \Dnut^{t-j} w(j) .	
\end{eqnarray*} 
Therefore, we have $x(t)=y(t) + \Dnut^{t-k} \Delta_{k-1} x(k-1)$, for $k \leq t < n$, where
\begin{eqnarray*}
	\Delta_{k-1} = D_{k-1}-\Dnut = B_0 \left(L_{k-1} - \Lmatrix{\para_0}\right).
\end{eqnarray*}
Thus, for we obtain 
\begin{eqnarray*}
	&& \instantcost{k-1}{\policy_k}- \instantcost{k-1}{\policy_{k-1}} \\
	&=&x(k-1)' \left( L_{k-1}' R L_{k-1} - \Lmatrix{\para_0}' R \Lmatrix{\para_0} \right) x(k-1).
\end{eqnarray*} 
Similarly, denote $P_0=Q+\Lmatrix{\para_0}' R \Lmatrix{\para_0}$, and replace for $x(t)$ to see that if $k \leq t <n$, then
\begin{eqnarray*}
	&&\instantcost{t}{\policy_k}-\instantcost{t}{\policy_{k-1}}\\
	&=&\left( 2 y(t)
	+ \Dnut^{t-k}\Delta_{k-1}x(k-1) \right)' P_0 \Dnut^{t-k} \Delta_{k-1} x(k-1).
\end{eqnarray*}
To proceed, plug-in for $y(t)$ to get \\$\instantcost{t}{\policy_k}-\instantcost{t}{\policy_{k-1}}= x(k-1)' F_{k-1}(t) x(k-1) + \eta_{k-1}(t)$, where $\Delta_{k-1}=D_{k-1}-\Dnut$ leads to
\begin{eqnarray*}
	\eta_{k-1}(t) &=& 2 x(k-1)' \Delta_{k-1}' {\Dnut'}^{t-k}P_0 \sum\limits_{j=k}^t \Dnut^{t-j} w(j), \\
	F_{k-1}(t) &=& D_{k-1}' {\Dnut'}^{t-k}P_0 \Dnut^{t-k} D_{k-1} -  {\Dnut'}^{t-k+1}P_0 \Dnut^{t-k+1}.
\end{eqnarray*}
Next, letting $z_k=\sum\limits_{t=k}^{n-1} \eta_{k-1}(t)$, and \\$G_{k}= L_{k-1}' R L_{k-1} - \Lmatrix{\para_0}' R \Lmatrix{\para_0} + \sum\limits_{t=k}^{n-1} F_{k-1}(t)$,
\\clearly 
\begin{eqnarray} \label{regretdecompos1.5}
\sum\limits_{t=0}^{n-1} \left[\instantcost{t}{\policy_k}-\instantcost{t}{\policy_{k-1}}\right]=x(k-1)' G_{k} x(k-1) + z_{k}. \:\:\:\:\:\:
\end{eqnarray}

To proceed, for $0 \leq j \leq n$ let $K_j = \sum\limits_{\ell=n-j}^{\infty}  {\Dnut'}^{\ell} P_0 \Dnut^{\ell}$. So, 
\begin{eqnarray*}
	\sum\limits_{t=k}^{n-1} F_{k-1}(t)= D_{k-1}' \left(K_n - K_k \right) D_{k-1} - \Dnut' \left(K_n - K_k \right) \Dnut
\end{eqnarray*}
implies $G_{k} = E_k + H_k$, where 
\begin{eqnarray*}
	E_k &=& - D_{k-1}' K_k D_{k-1} + \Dnut' K_k \Dnut,\\
	H_k &=&  L_{k-1}' R L_{k-1} - \Lmatrix{\para_0}' R \Lmatrix{\para_0} \\
	&-& \Dnut' K_n \Dnut+ D_{k-1}' K_n D_{k-1}.
\end{eqnarray*}
The Lyapunov equation (see~\cite{faradonbeh2018stabilization})
\begin{eqnarray} \label{lyapunoveq}
\Kmatrix{\para_0}- \Dnut' \Kmatrix{\para_0} \Dnut = P_0,
\end{eqnarray} 
leads to $K_n=\Kmatrix{\para_0}$. Thus, letting $X = L_{k-1} - \Lmatrix{\para_0}$, $ M = B_0'\Kmatrix{\para_0}B_0+R$, since $M\Lmatrix{\para_0}= - B_0' \Kmatrix{\para_0} A_0$, after doing some algebra we get
\begin{eqnarray*}
	H_k &=& \Lmatrix{\para_0}' R X + X' R \Lmatrix{\para_0} + \Dnut' \Kmatrix{\para_0} B_0 X \\
	&+& X' R X + X' B_0' \Kmatrix{\para_0} \Dnut + X' B_0' \Kmatrix{\para_0} B_0 X \\
	&=& X'MX
\end{eqnarray*}
Hence, adding up the terms in~\eqref{regretdecompos1.5}, \eqref{regretdecompos1} implies that
\begin{eqnarray} \label{regretdecompos2}
\regret{n}{\policy} = Z_n+ S_n+ T_n ,
\end{eqnarray}	 
where $Z_n = \sum\limits_{k=1}^{n}z_k$, $S_n = \sum\limits_{k=0}^{n-1} x(k)'E_{k+1}x(k)$, and $T_n = \sum\limits_{k=0}^{n-1} \norm{M^{1/2}\left(L_{k}-\Lmatrix{\para_0}\right) x(k)}{}^2$. In order to investigate $S_n$, we use the dynamics $x(k)=D_{k-1}x(k-1)+ w(k)$, as well as $\Dnut' K_{k+1} \Dnut = K_{k}$, to get 
\begin{eqnarray*}
	x(k)' \Dnut' &K_{k+1}& \Dnut x(k) = x(k-1)' D_{k-1}' K_{k} D_{k-1} x(k-1) \\
	&+& w(k)' K_{k} w(k) + 2 w(k)' K_{k} D_{k-1} x(k-1),
\end{eqnarray*}
for $0 < k < n$. Substituting in the expression for $S_n$, and denoting $w(0)=x(0)$, the telescopic differences vanish: 
\begin{eqnarray} 
&& S_n + x(n)' \Kmatrix{\para_0} x(n) \notag \\
&=& \sum\limits_{k=0}^{n-1} 2 w(k+1)' K_{k+1} D_{k} x(k)+ \sum\limits_{k=0}^{n} w(k)' K_{k} w(k). \label{Sdecompos}
\end{eqnarray}
Plugging
\begin{eqnarray*}
	D_kx(k)= \sum\limits_{j=0}^k \left( \Dnut^{j+1} w(k-j) + \Dnut^j \Delta_{k-j} x(k-j) \right),
\end{eqnarray*}
as well as $\optstate{n}=\sum\limits_{j=0}^n \Dnut^{n-j}w(j)$, in~\eqref{Sdecompos}, we have $\widetilde{S}_n=S_n + x(n)' K_n x(n) - \optstate{n}' K_n \optstate{n} = \sum\limits_{k=1}^{n} w(k)' K_{k} \xi_{k}$, where 
$\xi_{k} = 2 \sum\limits_{\ell=1}^{k} \Dnut^{\ell-1} \Delta_{k-\ell} x(k-\ell)$. Moreover, it is straightforward to show that $Z_n = \sum\limits_{j=1}^{n-1} \zeta_j' w(j)$, where 
\\$\zeta_j = 2 \sum\limits_{\ell=1}^{j} \sum\limits_{t=j}^{n-1}  {\Dnut'}^{t-j} P_0 \Dnut^{t-\ell} \Delta_{\ell-1} x(\ell-1)$. 
\\Hence, $\zeta_j = \left(K_n - K_j \right) \xi_j$ implies $\widetilde{S}_n+Z_n = \sum\limits_{k=1}^{n} w(k)' K_n \xi_{k}$. Next, we use the following lemma.

\begin{lemm} \cite{lai1982least} \label{Lailemma2}
	Suppose that for all $t \geq 0$, $y(t+1),v(t)$ are $\mathcal{G}_t$ measurable, $\mathcal{G}_t \subseteq \mathcal{G}_{t+1}$, and $\E{v(t+1) | \mathcal{G}_{t}}=0$. Define the martingale $\psi_n = \sum\limits_{t=1}^{n} y(t)'v(t)$, and let $\varphi_n = \sum\limits_{t=1}^{n} \norm{y(t)}{}^2$. If $\sup\limits_{t \geq 0} \E{\norm{v(t+1)}{}^2 \Big| \mathcal{G}_t} < \infty$, then 
	\begin{eqnarray*}
		\limsup\limits_{n \to \infty}\left| \psi_n \right| < \infty &\:\:\:\:\text{on}\:\:\:\:& \varphi_\infty < \infty, \\
		\limsup\limits_{n \to \infty} \frac{\psi_n}{\varphi_n^{1/2} \log \varphi_n} = 0 &\:\:\:\:\text{on}\:\:\:\:& \varphi_\infty = \infty.
	\end{eqnarray*}
\end{lemm}
Taking $\mathcal{G}_t = \sigma \left( \left\{ w(i) \right\}_{i=1}^t , \left\{ x(i) \right\}_{i=0}^t \right)$, and $v(t)=w(t)$, $y(t)=\xi_t$, we can use Lemma~\ref{Lailemma2} since Assumption~\ref{momentcondition} holds. So, stability of $\Dnut$ (Proposition~\ref{stabilizable}), and $\eigmin{M}>0$, lead to $\sum\limits_{k=1}^n \norm{\xi_k}{}^2 = \order{T_n}$. Thus, by~\eqref{regretdecompos2}, we get the desired result since $\widetilde{S}_n+Z_n = \order{T_n^{1/2} \log T_n}$.

Next, the first statement in Corollary~\ref{RegretCorollary} follows from Theorem 1 in the work of Lai and Wei~\cite{lai1985asymptotic}. To prove the second result, first observe that $S_\infty=S_n$, $T_\infty=T_n$, and $Z_\infty=Z_n$. Furthermore, note that for $t \geq n$ we have $\instantcost{t}{\policy} = x(t)' P_0 x(t)$, $\instantcost{t}{\optimalpolicy}= \optstate{t}' P_0 \optstate{t}$, as well as
\begin{eqnarray*}
	x(t) &=& \Dnut^{t-n} x(n)+ \sum\limits_{j=n+1}^t \Dnut^{t-j} w(j),\\ 
	\optstate{t} &=& \Dnut^{t-n} \optstate{n}+ \sum\limits_{j=n+1}^t \Dnut^{t-j} w(j).
\end{eqnarray*}
So, letting 
\begin{eqnarray*}
	\delta_n=2\sum\limits_{t=n}^\infty \sum\limits_{j=n+1}^t  \left(x(n)-\optstate{n}\right)' {\Dnut'}^{t-n} P_0 \Dnut^{t-j} w(j),
\end{eqnarray*}
by~\eqref{lyapunoveq} the following holds:
\begin{eqnarray*}
	\sum\limits_{t=n}^{\infty} \left[\instantcost{t}{\policy}-\instantcost{t}{\optimalpolicy}\right] = x(n)'K_n x(n)-\optstate{n}'K_n \optstate{n}+ \delta_n.
\end{eqnarray*}
Finally, \begin{eqnarray} \label{fluctuationmagnitude}
\norm{x(n)-\optstate{n}}{}^2= \norm{\sum\limits_{j=0}^{n-1} \Dnut^{n-1-j} \Delta_{j} x(j)}{}^2 = \order{T_n}, \:\:\:\:\:\:
\end{eqnarray}
together with Lemma~\ref{Lailemma2} imply $\delta_n=\order{T_n^{1/2} \log T_n}$.

\subsection{\bf Proof of Theorem \ref{adaptiveconsistency}}

	First, for an arbitrary $\para \in \mathcal{P}_0$, since $\para \in \nullspace{\para_0}$, we have 
	\begin{eqnarray} \label{eqnullspace}
	A+B \Lmatrix{\para_0} = A_0+ B_0 \Lmatrix{\para_0} = D_0.
	\end{eqnarray}
	Next, for an arbitrary fixed unit matrix (in the Frobenius norm) $X \in \R^{r \times p}$, let $L = \Lmatrix{\para_0} + \epsilon X$ be a linear feedback matrix which stabilizes the system of dynamics parameters $\para$. Note that according to Proposition \ref{stabilizable}, $\para \in \levelset{\para_0}$ leads to $\eigmax{\para \extendedLmatrix{\para_0}} <1$. Thus, $\eigmax{A+BL}<1$, as long as $\epsilon$ is sufficiently small. 
	
	Then, applying $L$ to the system $\para$, we get $\avecost{\para}{L} = \tr{ P \left( \epsilon \right) C }$, where $P \left( \epsilon \right)$ is the unique solution of the Lyapunov equation
	\begin{eqnarray} \label{generallyapunoveq}
	P \left( \epsilon \right) - \left( A+B L \right)' P \left( \epsilon \right) \left( A+B L \right) = Q + L'R L.
	\end{eqnarray}
	Note that according to \eqref{lyapunoveq} and \eqref{eqnullspace}, it holds that $P \left( 0 \right) = \Kmatrix{\para_0}$. Letting $\Delta \left( X \right) = \lim\limits_{\epsilon \to 0} {\epsilon}^{-1} \left({ P \left( \epsilon \right) - P \left( 0 \right)}\right)$,
	\eqref{generallyapunoveq} leads to 
	\begin{eqnarray} \label{derivativelyapunov}
	\Delta \left( X \right) - D_0' \Delta \left( X \right) D_0= X'N+ N'X,
	\end{eqnarray} 
	where $N=R \Lmatrix{\para_0}+ B' \Kmatrix{\para_0} D_0$. Next, $\para \in \levelset{\para_0}$ implies that $\Lmatrix{\para_0}$ is an optimal linear feedback for the system of dynamics parameter $\para$. So, the directional derivative of $\avecost{\para}{L}$ with respect to $L$ is zero in all directions. In the direction of $X$, the derivative is $\tr{ \Delta \left( X \right) C}$. Since all above statements hold regardless of the positive definite matrix $C$, \eqref{derivativelyapunov} and  $\tr{ \Delta \left( X \right) C}=0$ imply $N=0$;
	\begin{eqnarray} \label{conditionconsistency1}
	D_0' \Kmatrix{\para_0} B = - \Lmatrix{\para_0}' R.	
	\end{eqnarray}
	Therefore, \eqref{conditionconsistency1} is a necessary condition for $\para \in \mathcal{P}_0$. Note that according to \eqref{lyapunoveq} and \eqref{eqnullspace}, the necessary condition \eqref{conditionconsistency1} implies the necessity of $D_0' \Kmatrix{\para_0} A = \Kmatrix{\para_0} - Q$. 
	Further, for every input matrix $B$ which satisfies \eqref{conditionconsistency1}, the transition matrix $A$ will be uniquely determined by \eqref{eqnullspace} as $A=D_0-B\Lmatrix{\para_0}$.
	
	Conversely, suppose that $B$ is an arbitrary matrix which satisfies \eqref{conditionconsistency1}. Letting $A=D_0-B\Lmatrix{\para_0}$, we show that $\left[A,B\right] = \para \in \mathcal{P}_0$. For this purpose, since the above definition of $A$ automatically leads to $\para \in \nullspace{\para_0}$, it suffices to show $\para \in \levelset{\para_0}$. Writing $Y = B-B_0$, we get $A=A_0-Y \Lmatrix{\para_0}$. Moreover, define $G = A' \Kmatrix{\para_0} A$, $H = B' \Kmatrix{\para_0} A$, $M = B_0' \Kmatrix{\para_0} B_0 +R$, and $S = B_0' \Kmatrix{\para_0}Y + Y' \Kmatrix{\para_0} B_0 + Y' \Kmatrix{\para_0} Y$. Then, we calculate the matrix
	\begin{eqnarray*}
		V &=& Q + G - H' \left( M+S \right)^{-1} H = Q + A'\Kmatrix{\para_0}A \\
		&-& A' \Kmatrix{\para_0}B \left(B'\Kmatrix{\para_0}B+R\right)^{-1} B'\Kmatrix{\para_0}A. 
	\end{eqnarray*}
Writing $A,B,G,H$ in terms of $A_0,B_0,M,S,Y$, we have
\begin{eqnarray*}
	V = Q &+& A_0' \Kmatrix{\para_0} A_0 + \Lmatrix{\para_0}' S \Lmatrix{\para_0} - \big[ B_0' \Kmatrix{\para_0} A_0 \\
	&-& S \Lmatrix{\para_0} \big]' \left( M+S \right)^{-1} \big[ B_0' \Kmatrix{\para_0} A_0 - S \Lmatrix{\para_0} \big]
\end{eqnarray*}
Then, using $\left( M+S \right)^{-1} = M^{-1} - \left( M+S \right)^{-1} S M^{-1}$, \eqref{ricatti2}, and $M \Lmatrix{\para_0} = -B_0' \Kmatrix{\para_0} A_0$, $V$ can be written as $V = \Kmatrix{\para_0} + \Lmatrix{\para_0}' S W$, where
\begin{eqnarray*}
W &=& \Lmatrix{\para_0} 
- \left( M+S \right)^{-1} \left(S \Lmatrix{\para_0} 
+ B_0' \Kmatrix{\para_0} A_0\right) \\
&=& \Lmatrix{\para_0} - \left( M+S \right)^{-1} \left(S +M \right) \Lmatrix{\para_0} = 0 ;
\end{eqnarray*}
i.e., $V= \Kmatrix{\para_0}$ is a solution of the Riccati equation \eqref{ricatti2} for $\para$. According to Proposition \ref{stabilizable}, the solution is unique; which is $\Kmatrix{\para} = \Kmatrix{\para_0}$. Moreover, $\Lmatrix{\para} = - \left( M + S \right)^{-1} H = \Lmatrix{\para_0}$ 
	shows that $\para \in \levelset{\para_0}$. So far, we have shown that $\para \in \mathcal{P}_0$, if and only if \eqref{eqnullspace} and \eqref{conditionconsistency1} hold. Next, \eqref{conditionconsistency1} is essentially stating that every column of $B-B_0$ (which is a vector in $\R^p$), is orthogonal to the all columns of $\Kmatrix{\para_0} D_0$. This verifies that \eqref{conditionconsistency1} specifies a shifted linear subspace. To find the dimension, since $B$ has $r$ columns, and \eqref{eqnullspace} uniquely determines $A$ in terms of $B$, we get $\dimension{}{\mathcal{P}_0} = \left( p - \rank{\Kmatrix{\para_0} D_0}\right)r$.
	Finally, by positive definiteness of $Q$, \eqref{lyapunoveq} implies $\rank{\Kmatrix{\para_0}}=p$. Further, since $D_0= \left[ I_p - B_0 M^{-1} B_0' \Kmatrix{\para_0} \right] A_0$,
	it suffices to show
	\begin{eqnarray} \label{rankequivalence}
	\rank{I_p - B_0 M^{-1} B_0' \Kmatrix{\para_0}}=p.
	\end{eqnarray}
	If \eqref{rankequivalence} does not hold, there exists $v \in \R^p$ such that $v \neq 0$ and $v = B_0 M^{-1} B_0' \Kmatrix{\para_0} v$. So, $v= B_0 \widetilde{v}$ where $\widetilde{v} = M^{-1} B_0' \Kmatrix{\para_0} v \in \R^r$. Thus,
	\begin{eqnarray*}
	B_0' \Kmatrix{\para_0} B_0 \widetilde{v} = B_0' \Kmatrix{\para_0} v = M \widetilde{v}
	= \left[ B_0' \Kmatrix{\para_0} B_0 + R \right] \widetilde{v},
	\end{eqnarray*} 
	or equivalently, $R \widetilde{v}=0$. Positive definiteness of $R$ implies that $\widetilde{v}=0$, which contradicts $B_0 \widetilde{v} \neq 0$. This proves \eqref{rankequivalence}, which completes the proof.

\subsection{\bf Proof of Theorem \ref{RCETheorem}}

	The proof is based on a sequence of intermediate results. First, for $i\geq 1$, let $V_i$ be the (unnormalized) state covariance during the $i$-th episode: $V_i = \sum\limits_{t=\lfloor \rrate^{i-1} \rfloor}^{\lfloor \rrate^{i} \rfloor -1} x(t) x(t)'$.
	\begin{lemm} \label{smallcov}
		For the matrix $V_i$ defined above, the followings hold: $\eigmax{V_m} = \order{\rrate^m}$, $\liminf\limits_{m \to \infty} \rrate^{-m} \eigmin{V_m} \geq \left( \rrate - 1 \right) \eigmin{C}$.
	\end{lemm}
	Then, in order to study the behavior of the least-squares estimate in \eqref{RCEalgo}, define
	\begin{eqnarray*}
		U_i = \sum\limits_{t=0}^{\lfloor \rrate^{i} \rfloor -1} \extendedLmatrix{\estpara{t}} x(t) x(t)' \extendedLmatrix{\estpara{t}}'.
	\end{eqnarray*}
	Note that since the parameter $\estpara{t}$ remains set (not changing) during each episode, $U_i$ can be written in terms of $V_1, \cdots, V_i$ as follows. First, for all $\lfloor \rrate^{i-1} \rfloor \leq t \leq \lfloor \rrate^{i} \rfloor -1$, the parameter estimate $\estpara{t}$ does not change. So, if $t$ belongs to the $i$-th episode, define the linear feedback matrix is $L_i= \Lmatrix{\estpara{t}}$. Letting $\widetilde{L}_{i}=\extendedLmatrix{\estpara{t}}$, we have $U_i = \sum\limits_{j=1}^i \widetilde{L}_{j} V_j \widetilde{L}_{j}'$. Then, the smallest eigenvalue of $U_i$ follows a different lower bound compared to that of $V_i$:
	\begin{lemm} \label{bigcov}
		Define $U_m$ as above. Then, we have $\liminf\limits_{m \to \infty} {\rrate^{-m/2}} \eigmin{U_m} >0 $, and $\eigmax{U_m} = \order{\rrate^{m}}$.
	\end{lemm}
	Next, the following result states that the estimation accuracy is determined by the eigenvalues of $U_i$. 
	\begin{lemm} \cite{lai1982least} \label{Lailemma}
		For $n = \lfloor \rrate^m \rfloor$, define $\estpara{n}, \optpara{n}$ according to \eqref{RCEalgo}. Then, we have
		\begin{eqnarray*}
			\Mnorm{\estpara{n} - \optpara{n} - \para_0}{}^2 = \order{ \frac{\log \eigmax{U_m}}{\eigmin{U_m}}}. 
		\end{eqnarray*}
	\end{lemm}
	Therefore, Lemma \ref{bigcov} leads to $\Mnorm{\estpara{n}-\optpara{n}-\para_0}{}= \order{n^{-1/4} \log^{1/2}n }$. Using the moment condition in Remark \ref{perturbmomentcondition}, Markov's inequality gives $\PP{\Mnorm{\randommatrix_m}{}> m^{1/4}} = \order{m^{-1-\epsilon/4}}$. Thus, an application of the Borel-Cantelli Lemma leads to $\Mnorm{\randommatrix_m}{}= \order{m^{1/4}}$; i.e., $\Mnorm{\optpara{n}}{} = \order{n^{-1/4} \log^{1/2}n }$. So, we get the desired result about the identification rate: $\Mnorm{\estpara{n}-\para_0}{}= \order{n^{-1/4} \log^{1/2}n }$. To proceed, we present the following auxiliary result which shows that a similar rate holds for the deviations from the optimal linear feedback.
	\begin{lemm} \cite{faradonbeh2017finite} \label{Lipschitzlemma}
		There exist $0<\epsilon_0,\Lipschitz{L}<\infty$, such that for all stabilizable $\para$ satisfying $\Mnorm{\para-\para_0}{}< \epsilon_0$, the following holds: $\Mnorm{\Lmatrix{\para}-\Lmatrix{\para_0}}{} \leq \Lipschitz{L} \Mnorm{\para-\para_0}{}$.
	\end{lemm}
	So, utilizing Lemma~\ref{Lipschitzlemma}, we have
	\begin{eqnarray} \label{randomaccuracy}
	\Mnorm{\Lmatrix{\estpara{n}}-\Lmatrix{\para_0}}{} = \order{ \frac{\log^{1/2}n}{n^{1/4}} + \Mnorm{\optpara{n}}{}}.
	\end{eqnarray}
	On the other hand, since the policy is not being updated during each episode, we can write down the regret in terms of the matrices $V_i$. Henceforth in the proof, suppose that the time $n$ belongs to the $m$-th episode: $\lfloor \rrate^{m-1} \rfloor \leq n < \lfloor \rrate^m \rfloor$. Then, applying Theorem~\ref{RegretTheorem} and Corollary~\ref{RegretCorollary}, we get
	\begin{eqnarray*}
		\regret{n}{\adaptivepolicy} &=& \order{\sum\limits_{i=0}^{m} \left(L_i - \Lmatrix{\para_0} \right) V_i \left(L_i - \Lmatrix{\para_0} \right)' + \rrate^{m/2}}  \\
		&=& \order{\sum\limits_{i=0}^{m} \rrate^i \Mnorm{L_i-\Lmatrix{\para_0}}{}^2 +\rrate^{m/2}}, 
	\end{eqnarray*}
	where in the last equality above we applied Lemma \ref{smallcov}. Based on the definition of the perturbation $\optpara{n}$ in terms of the random matrix $\randommatrix_m$, define
	\begin{eqnarray*}
		S_m = \sum\limits_{i=0}^{m} i^{1/2}\rrate^{i/2} \Mnorm{\randommatrix_i}{}^2, \:\:\:\:\:\:
		T_m = \sum\limits_{i=0}^{m} i^{3/4}\rrate^{i/2} \Mnorm{\randommatrix_i}{} .
	\end{eqnarray*}
	So, by \eqref{randomaccuracy}, the regret is in magnitude dominated by $S_m,T_m$, and $m \rrate^m/2$: $\regret{n}{\adaptivepolicy} = \order{S_m+T_m+ m \rrate^{m/2}}$. Note that as $m$ and $n$ grow, the magnitudes of $n^{1/2} \log n$ and $m \rrate^{m/2}$ is the same. Finally, the following lemma leads to the desired result:
	\begin{lemm} \label{stochterm}
		For the terms $S_m,T_m$ defined above the followings hold: $S_m = \order{m \rrate^{m/2}}$, $T_m = \order{m \rrate^{m/2}}$.
	\end{lemm}

\subsection{\bf Proof of Theorem \ref{TSTheorem}}

	In this proof, we use the following result. 
	\begin{lemm} \label{TScov}
		For the matrix $\Sigma_m$ defined in \eqref{TSalgo2} we have $\liminf\limits_{m \to \infty} \rrate^{-m/2} m^{1/2} \eigmin{\Sigma_m} >0$, $\eigmax{\Sigma_m} = \order{\rrate^{m}}$.
	\end{lemm}
	Hence, since $\mu_m$ is the least-squares estimate, and $\Sigma_m$ is the unnormalized empirical covariance matrix, Lemma~\ref{Lailemma} leads to $\Mnorm{\mu_m-\para_0}{}= \order{\rrate^{-m/4} m^{} }$. Then, because every row of $\estpara{\lfloor \rrate^m \rfloor}-\mu_m$ is a mean zero Gaussian with covariance matrix $\Sigma_m^{-1}$, by Lemma \ref{TScov} we have
	\begin{eqnarray*}
		\sum\limits_{m=0}^\infty \PP{\Mnorm{\estpara{\lfloor \rrate^m \rfloor}-\mu_m}{}> \rrate^{-m/4} m^{}} < \infty.
	\end{eqnarray*}
	Thus, Borel-Cantelli Lemma leads to the desired result about the identification rate: $\Mnorm{\estpara{\lfloor \rrate^m \rfloor}-\para_0}{}= \order{\rrate^{-m/4} m^{} }$. By Lemma \ref{Lipschitzlemma}, a similar rate holds for the linear feedbacks: $\Mnorm{\Lmatrix{\estpara{\lfloor \rrate^m \rfloor}}-\Lmatrix{\para_0}}{}= \order{\rrate^{-m/4} m^{} }$. Finally, plugging in the expression of Theorem~\ref{RegretTheorem}, and utilizing Corollary~\ref{RegretCorollary}, we get the desired result for the regret:
	\begin{eqnarray*}
		\regret{\lfloor \rrate^m \rfloor}{\adaptivepolicy} &=& \order{\sum\limits_{i=0}^{m} \rrate^i \Mnorm{\Lmatrix{\estpara{\lfloor \rrate^m \rfloor }}-\Lmatrix{\para_0}}{}^2 + \rrate^{m/2}} \\
		&=& \order{\sum\limits_{i=0}^{m} \rrate^{i/2} i^2}
		= \order{ \rrate^{m/2} m^2} . 
	\end{eqnarray*}

\subsection{\bf Proof of Theorem \ref{GCETheorem}}

	Define $V_i,U_i,L_i,\widetilde{L}_i$ similar to the proof of Theorem \ref{RCETheorem}. Further, for $i\geq 1$, let $n_i= \lfloor \rrate^i \rfloor -1 $ be the end time of episode $i$, and denote 
	\\$\loss{i}{}{\para} = \sum\limits_{t=0}^{n_i-1} \norm{x(t+1)- \para \extendedLmatrix{\estpara{t}} x(t)}{}^2$.
	\\Letting $\para_\star= \arg\min\limits_{\para \in \R^{p \times q}} \loss{i}{}{\para}$ for a fixed $i$, it is straightforward to show that 
	\begin{eqnarray*}
		\loss{i}{}{\para}= \tr{\left( \para- \para_\star\right) U_i \left( \para- \para_\star\right)' } - \tr{ \para_\star U_i \para_\star'}.
	\end{eqnarray*}
	Therefore, since $\para_0 \in \tempparaspace{0}$, \eqref{GCEalgo} implies that $\loss{i}{}{\estpara{n_i}-\optpara{n_i}} \leq \loss{i}{}{\para_0}$. So, the triangle inequality leads to
	\begin{eqnarray*}
		&& \tr{\left( \estpara{n_i}-\optpara{n_i}-\para_0 \right) U_i \left( \estpara{n_i}-\optpara{n_i}-\para_0 \right)' }\\ &\leq& 4\tr{\left( \para_\star-\para_0 \right) U_i \left( \para_\star-\para_0 \right)' }
	\end{eqnarray*}
	Hence, the normal equation $\left( \para_\star-\para_0 \right) U_i =  \sum\limits_{t=0}^{n_i-1} w(t+1) x(t)'\extendedLmatrix{\estpara{t}}'$, in addition to Lemma~\ref{bigcov} and Lemma~\ref{Lailemma} imply that
	\begin{eqnarray*}
		\tr{\left( \estpara{n_i}-\optpara{n_i}-\para_0 \right) U_i \left( \estpara{n_i}-\optpara{n_i}-\para_0 \right)' } = \order{i}
	\end{eqnarray*}
	Applying Lemma \ref{smallcov}, we obtain
	\begin{eqnarray} \label{GCEproofeq1}
		\sum\limits_{j=0}^i \rrate^{j} \Mnorm{\left( \estpara{n_i}-\optpara{n_i} - \para_0 \right) \widetilde{L}_j}{}^2 = \order{ i } .
	\end{eqnarray}
	Since $\optpara{n_j}= \order{n_j^{-1/2}}$, by Lemma~\ref{Lipschitzlemma} we have $\Mnorm{L_j - \Lmatrix{\estpara{n_j}-\optpara{n_j}}}{} = \order{\rrate^{-j/2}}$. Hence, 
	\begin{eqnarray*}
		\sum\limits_{j=0}^i \rrate^{j} \Mnorm{\left( \estpara{n_i} - \optpara{n_i} - \para_0 \right) \extendedLmatrix{\estpara{n_j}-\optpara{n_j}}}{}^2 = \order{ i} .
	\end{eqnarray*}
	Using $\estpara{n_j}-\optpara{n_j} \in \tempparaspace{0}$, \eqref{Sideq} leads to \\$\Mnorm{ \Lmatrix{\estpara{n_i} - \optpara{n_i}} - \Lmatrix{\para_0} }{} = \order{ i^{1/2}\rrate^{-i/2} }$, which by Lemma \ref{Lipschitzlemma} implies that
	\begin{eqnarray} \label{GCEproofeq2}
		\Mnorm{ \Lmatrix{\estpara{n_i}} - \Lmatrix{\para_0} }{} = \order{i^{1/2}\rrate^{-i/2} }.
	\end{eqnarray}
	Thus, we have 
	\begin{eqnarray} 
		\sum\limits_{t=0}^{n_m-1} \norm{ \left(\Lmatrix{\para_0} - L_t \right) x(t) }{}^2 &=& \order{\sum\limits_{i=0}^{m} \rrate^i \Mnorm{L_i-\Lmatrix{\para_0}}{}^2} \notag \\
		&=& \order{m^2}. \label{GCEdecompos1} 
	\end{eqnarray}
	Moreover, putting Assumption~\ref{momentcondition}, Corollary~\ref{RegretCorollary}, \eqref{fluctuationmagnitude}, and \eqref{GCEproofeq2} together, we obtain $\norm{x(n_m)-\optstate{n_m}}{} \norm{\optstate{n_m}}{} = \order{m}$, which in turn leads to
	\begin{eqnarray}\label{GCEdecompos2}
		\optstate{n_m}' \Kmatrix{\para_0} \optstate{n_m} - x(n_m)' \Kmatrix{\para_0} x(n_m) = \order{m}.\:\:\:\:\:\:\:\:\:\:\:\:
	\end{eqnarray}
	Then, \eqref{GCEdecompos1} and \eqref{GCEdecompos2} lead to the desired result for the regret: $\regret{n_m}{\adaptivepolicy} = \order{m^2}$. Further, \eqref{GCEproofeq1} and \eqref{GCEproofeq2} imply that
	\begin{eqnarray*}
	\Mnorm{\left( \estpara{n_m}-\optpara{n_m} - \para_0 \right) \extendedLmatrix{\para_0}}{} = \order{ \rrate^{-m/2} m^{1/2} };
	\end{eqnarray*}
	i.e., $\inf\limits_{\para \in \nullspace{\para_0}} \Mnorm{\estpara{n_m}-\para}{} = \order{\rrate^{-m/2} m^{1/2}}$. Finally, since \eqref{GCEproofeq2} implies a similar result for $\levelset{\para_0}$, the desired result for $\mathcal{P}_0$ holds.

\section{Proofs of Auxiliary Results}
\begin{pf*}{Proof of Proposition \ref{diffterm}}
	Under the optimal regulator $\optimalpolicy$ the closed-loop transition matrix is $D=A_0+B_0 \Lmatrix{\para_0}$. Denoting $P=Q+\Lmatrix{\para_0}'R\Lmatrix{\para_0}$, the instantaneous cost is $\instantcost{t}{\optimalpolicy}= x(t)'Px(t)$. So, by Proposition \ref{stabilizable} we have
	\begin{eqnarray*}
	\sum\limits_{t=0}^{n-1}x(t)'Px(t) - n \optcost{\para_0}=\tr{PV_n}- n\tr{\Kmatrix{\para_0}C},
	\end{eqnarray*}
	where $V_n= \sum_{t=0}^{n-1} x(t)x(t)'$. Then, define the following matrices:
	\begin{eqnarray*}
		U_n &=& \sum\limits_{t=0}^{n-1} \left[Dx(t)w(t+1)'+w(t+1)x(t)'D'\right] ,\\
		C_n &=& \sum\limits_{t=1}^{n} w(t) w(t)', \\
		E_n &=& U_n+C_n+ x(0)x(0)' - x(n)x(n)'.
	\end{eqnarray*}
	Using the dynamics equation $x(t+1)=Dx(t)+w(t+1)$, after doing some algebra we get the Lyapunov equation $V_{n}=DV_{n}D'+E_n$; i.e. $V_n = \sum\limits_{k=0}^\infty D^k E_n {D'}^k$. Using \eqref{lyapunoveq}, we can write
	\begin{eqnarray*}
		&& \tr{PV_n}- n\tr{\Kmatrix{\para_0}C} \\
		&=& \tr{\left(C_n - nC + U_n + x(0)x(0)' - x(n)x(n)'\right)\Kmatrix{\para_0}}.
	\end{eqnarray*}
	According to Corollary~\ref{RegretCorollary}, we have $\norm{x(0)}{}^2+\norm{x(n)}{}^2 = \order{n^{1/2}}$. Further, Lemma \ref{Lailemma2} implies that $U_n = \order{n^{1/2}\log n}$. Since the moment condition of Assumption~\ref{momentcondition} implies $\sup\limits_{t \geq 1} \E{\Mnorm{w(t)w(t)'-C}{}^2} < \infty$, applying Lemma \ref{Lailemma2} we get $C_n-nC = \order{n^{1/2} \log n}$, which completes the proof.
\end{pf*}
\begin{pf*}{Proof of Lemma \ref{unfalselemma}}
	Clearly, we can write
	\begin{eqnarray*}
		\equilib{\para_0} = \bigcup\limits_{k=0}^p \paraspace{k},
	\end{eqnarray*}	
	where 
	\begin{eqnarray*}
		\paraspace{k} = \left\{ \para \in \equilib{\para_0}: \rank{A}=k \right\} \in \R^{p \times q}.
	\end{eqnarray*}
	Then, for a fixed $0 \leq k \leq p$, suppose that $\para_1 \in \paraspace{k}$ is arbitrarily chosen. Note that $\para_1 \in \equilib{\para_0}$ is equivalent to $\para_0 \in \nullspace{\para_1}$. If there exists some $\para_2 \in \levelset{\para_1}$ such that $\para_2 \in \equilib{\para_0}$, then
	\begin{eqnarray*}
		\para_2 \extendedLmatrix{\para_1} = \para_2 \extendedLmatrix{\para_2} = \para_0 \extendedLmatrix{\para_2} = \para_0 \extendedLmatrix{\para_1} = \para_1 \extendedLmatrix{\para_1},
	\end{eqnarray*}
	i.e. $\para_2 \in \nullspace{\para_1}$. Therefore, according to \eqref{asymptoticuncertainty}, the matrix $\para_2$ belongs to the shifted linear subspace $\nullspace{\para_1} \cap \levelset{\para_1}$, and
	\begin{eqnarray} \label{unfalselemmaeq1}
	\dimension{}{\nullspace{\para_1} \cap \levelset{\para_1}} = \left( p - k \right) r.
	\end{eqnarray}
	Next, for $k=0,1, \cdots, p$, define 
	\begin{eqnarray*}
		\mathcal{Y}_k = \left\{ \Lmatrix{\para}: \para \in \paraspace{k} \right\} \subset \R^{r \times p}.
	\end{eqnarray*}
	For $\para_1 \in \paraspace{k}$, it holds that $\rank{A_1}=k$. Let the vectors $v_1, \cdots , v_{p-k} \in \R^p$ be such that $A_1 v_j=0$, for $1 \leq j \leq p-k$. Then, according to the definition of $\Lmatrix{\para_1}$ in \eqref{ricatti1}, we have $\Lmatrix{\para_1} v_j=0$, for $1 \leq j \leq p-k$. Hence, since every matrix $\Lmatrix{\para}$ has $r$ rows, we get
	\begin{eqnarray} \label{unfalselemmaeq2}
	\dimension{}{\mathcal{Y}_k} = kr.
	\end{eqnarray}
	To proceed, using 
	\begin{eqnarray*}
		\paraspace{k} = \bigcup\limits_{L \in \mathcal{Y}_k} \left\{ \para \in \equilib{\para_0} : \Lmatrix{\para}=L \right\},
	\end{eqnarray*}
	\eqref{unfalselemmaeq1}, \eqref{unfalselemmaeq2} imply
	\begin{eqnarray*}
		\dimension{}{\paraspace{k}} \leq \dimension{}{\mathcal{Y}_k} + \dimension{}{\nullspace{\para_1} \cap \levelset{\para_1}} = pr.
	\end{eqnarray*}
	So, $\dimension{}{\equilib{\para_0}}=pr$, which yields to the desired result.
\end{pf*}
\begin{pf*}{Proof of Lemma \ref{optmanifold}}
	For $\para \in \levelset{\para_0}$, let $\para=\para_0 + \epsilon \left[ M,N \right]$, where $M \in \R^{p \times p}, N \in \R^{p \times r}$. First, we calculate the matrix
	\begin{eqnarray*}
	\Delta = \lim\limits_{\epsilon \to 0} \frac{\Kmatrix{\para}-\Kmatrix{\para_0}}{\epsilon}.
	\end{eqnarray*}
	Define $D=\para\extendedLmatrix{\para}, D_0=\para_0 \extendedLmatrix{\para_0}$. Note that
	\begin{eqnarray*}
	\lim\limits_{\epsilon \to 0} \frac{D-D_0}{\epsilon} = M+N\Lmatrix{\para_0},
	\end{eqnarray*}
	since $\Lmatrix{\para}=\Lmatrix{\para_0}$. Further, according to \eqref{lyapunoveq}, $\Delta$ is the unique solution of the Lyapunov equation 
	\begin{eqnarray*}
	\Delta - D_0'\Delta D_0 = D_0'Z+Z'D_0,
	\end{eqnarray*}
	where $Z = \Kmatrix{\para_0} \left( M + N\Lmatrix{\para_0} \right)$. 
	
	Then, defining the matrices
	\begin{eqnarray*}
		X &=& B_0' \Delta A_0 + B_0' \Kmatrix{\para_0} M + N' \Kmatrix{\para_0} A_0, \\
		Y &=& B_0' \Delta B_0 + B_0' \Kmatrix{\para_0} N + N' \Kmatrix{\para_0} B_0, 
	\end{eqnarray*}
	the followings hold:
	\begin{eqnarray*}
		\lim\limits_{\epsilon \to 0} \frac{B'\Kmatrix{\para}A - B_0'\Kmatrix{\para_0}A_0 }{\epsilon} &=& X ,\\ \lim\limits_{\epsilon \to 0} \frac{B'\Kmatrix{\para}B - B_0'\Kmatrix{\para_0}B_0 }{\epsilon} &=& Y . 
	\end{eqnarray*}
	Using \eqref{ricatti1}, after doing some algebra we get $X+Y \Lmatrix{\para_0}=0$. Substituting for $X,Y$ it leads to
	\begin{eqnarray} \label{tangentspace}
	B_0' Z + \left(N' \Kmatrix{\para_0} + B_0' \Delta \right) D_0 =0.
	\end{eqnarray}
	Thus, the tangent space of $\levelset{\para_0}$ at point $\para_0$ consists of matrices $\left[ M,N \right]$ which satisfy \eqref{tangentspace}. Note that $\Delta$ is uniquely determined according to $Z$. 
	
	To find the dimension of solutions of \eqref{tangentspace}, first let $\mathcal{Z} \subset \R^{p \times p}$ be the set of matrices $Z$, such that the equation $B_0'Z=TD_0$ has a solution $T \in \R^{r \times p}$. Further, for $k= p -\rank{D_0}$, let $v_1, \cdots, v_k \in \R^{p}$ be orthonormal vectors satisfying $D_0v_i=0$. Putting the above vectors together, define the matrix $V= \left[ v_1, \cdots, v_k \right]$. Similarly, denote the orthonormal basis of the columns of $B_0$ by $b_1, \cdots, b_m$, where $m=\rank{B_0}$. Now, the equation $B_0'Z=TD_0$ has a solution if and only if $B_0'ZV=0$. So,
	\begin{eqnarray*}
	\mathcal{Z} &=& \left\{ Z \in \R^{p \times p}: B_0'ZV=0 \right\} \\
	&=& \left\{ Z \in \R^{p \times p}: \tr{Zv_ib_j'}=0, \forall 1 \leq i \leq k , \forall 1 \leq j \leq m \right\}.
	\end{eqnarray*}
	Note that $\tr{\cdot}$ is an inner product on the set of $p \times p$ matrices. Moreover, all matrices $v_ib_j'$, $1 \leq i \leq k$, $1 \leq j \leq m$ are orthogonal, and so linearly independent. To see that, calculating the inner products, as long as $i_1 \neq i_2$ or $j_1 \neq j_2$, we have
	\begin{eqnarray*}
	\tr{b_{j_1}v_{i_1}'v_{i_2}b_{j_2}'} &=& v_{i_1}'v_{i_2} b_{j_2}'b_{j_1} = 0.
	\end{eqnarray*} 
	Therefore,
	\begin{eqnarray} \label{dimensioneq1}
	\dimension{}{\mathcal{Z}} = p^2 - \left( p - \rank{D_0} \right) \rank{B_0}.
	\end{eqnarray}
	Similar to the proof of Theorem \ref{adaptiveconsistency}, for any fixed matrix $Z \in \mathcal{Z}$, the set of matrices $N$ satisfying \eqref{tangentspace} is of dimension
	\begin{eqnarray} \label{dimensioneq2}
	\left( p - \rank{D_0} \right) r.
	\end{eqnarray}
	Note that since $\Kmatrix{\para_0}$ is invertible, every pair $Z,N$ uniquely determines the matrix $M$. Putting \eqref{dimensioneq1} and \eqref{dimensioneq2} together, the desired result is implied since $\rank{D_0}=\rank{A_0}$ (see the proof of Theorem \ref{adaptiveconsistency}).
\end{pf*}

\begin{pf*}{Proof of Lemma \ref{smallcov}}
	First, once the system is stabilized, we have $x(t)=D_i x(t-1)+ w(t)$, where of $D_i =\para_0 \widetilde{L}_i$ is the stable closed-loop matrix during the $i$-th episode. Thus, 
	\begin{eqnarray*}
		V_i &=& \sum\limits_{t=\lfloor \rrate^{i-1} \rfloor}^{\lfloor \rrate^{i} \rfloor -1} x(t) x(t)'\\
		&=& x(\lfloor \rrate^{i-1} \rfloor) x(\lfloor \rrate^{i-1} \rfloor)' - x(\lfloor \rrate^{i} \rfloor) x(\lfloor \rrate^{i} \rfloor)' \\
		&+& \sum\limits_{t=\lfloor \rrate^{i-1} \rfloor}^{\lfloor \rrate^{i} \rfloor -1} \left(D_i x(t)+w(t+1)\right) \left(D_i x(t)+w(t+1)\right)' \\
		&=& D_i V_i D_i' + C_i + E_i +F_i,
	\end{eqnarray*}
	where 
	\begin{eqnarray*}
		C_i &=& \sum\limits_{t=\lfloor \rrate^{i-1} \rfloor}^{\lfloor \rrate^{i} \rfloor -1} w(t+1) w(t+1)', \\
		E_i &=& \sum\limits_{t=\lfloor \rrate^{i-1} \rfloor}^{\lfloor \rrate^{i} \rfloor -1} D_i x(t) w(t+1)'+ w(t+1) x(t)'D_i', \\
		F_i &=& x(\lfloor \rrate^{i-1} \rfloor) x(\lfloor \rrate^{i-1} \rfloor)' - x(\lfloor \rrate^{i} \rfloor) x(\lfloor \rrate^{i} \rfloor)'.
	\end{eqnarray*}
	Then, by the Law of Large Numbers, Assumption \ref{momentcondition} implies that
	\begin{eqnarray} \label{auxeq1}
	\lim\limits_{m \to \infty} \rrate^{-m+1} C_m = \left( \rrate -1 \right) C. 
	\end{eqnarray}
	In addition, by the Martingale Convergence Theorem,
	\begin{eqnarray} \label{auxeq2}
	\limsup\limits_{m \to \infty} \rrate^{-m} \Mnorm{E_m}{} = 0. 
	\end{eqnarray}
	Finally, since the system is stable in the average sense, similar to Corollary~\ref{RegretCorollary} we have 
	\begin{eqnarray} \label{auxeq3}
	\limsup\limits_{m \to \infty} \rrate^{-m} \Mnorm{F_m}{} = 0. 
	\end{eqnarray}
	Putting \eqref{auxeq1}, \eqref{auxeq2}, and \eqref{auxeq3} together, the Lyapunov equation $V_m = D_m V_m D_m' + C_m + E_m +F_m$ has the solution
	\begin{eqnarray*}
		\lim\limits_{m \to \infty} \rrate^{-m+1} V_m = \left( \rrate -1 \right) \lim\limits_{m \to \infty} \sum\limits_{k=0}^\infty D_m ^k C {D_m'}^k.
	\end{eqnarray*}
	By stability of $D_m$, the RHS of the above equation is $\order{1}$; i.e. $\eigmax{V_m}=\order{\rrate^m}$. Moreover,
	\begin{eqnarray*}
		\eigmin{\sum\limits_{k=0}^\infty D_m ^k C {D_m'}^k} \geq \eigmin{C}
	\end{eqnarray*} 
	leads to the desired result about the smallest eigenvalue of $V_m$.
\end{pf*}
\begin{pf*}{Proof of Lemma \ref{bigcov}}
	First, Lemma \ref{smallcov} implies that $\eigmax{U_m} = \order{\rrate^m}$. To show the desired result on the smallest eigenvalue of $U_m$, let $v \in \R^{q}$ be an arbitrary unit vector ($\norm{v}{}=1$). Then, for $i=1,\cdots, m$, define the $p$ dimensional vectors $z_{i} = \rrate^{i/4} \widetilde{L}_i' v$. Using Lemma \ref{smallcov} we get
	\begin{eqnarray*}
		&& \rrate^{-m/2} v' U_m v \\
		&\geq& \sum\limits_{i= \lfloor m/2 \rfloor}^m \rrate^{-m/2-i/2} z_{i}' V_i z_{i} \\
		&\geq& \left( \rrate - 1 \right) \eigmin{C} \sum\limits_{i= \lfloor m/2 \rfloor}^m \rrate^{-m/2+i/2} \norm{z_{i}}{}^2\\
		&\geq& \left( \rrate - 1 \right) \eigmin{C} \rrate^{-k/2} \sum\limits_{i= m-k}^m \norm{z_{i}}{}^2,		
	\end{eqnarray*}
	where $k$ is large enough to satisfy $ kp \geq q +4$. Next, define the $(k+1)p \times q$ matrix
	\begin{eqnarray} \label{keymatrixeq}
		M_m = \begin{bmatrix}
			\rrate^{(m-k)/4} I_p \:\:\:\:\:\:\:& \rrate^{(m-k)/4} L_{m-k}'\\
			\vdots & \vdots \\
			\rrate^{(m-1)/4} I_p & \rrate^{(m-1)/4} L_{m-1}' \\
			\rrate^{m/4} I_p & \rrate^{m/4} L_m' 
		\end{bmatrix}.
	\end{eqnarray}
	On the event $\eigmin{U_m} \neq \orderinv{\rrate^{m/2}}$, we have:
	\begin{eqnarray*}
		\liminf\limits_{m \to \infty} \sum\limits_{i= m-k}^m \norm{z_{i}}{}^2 =0,
	\end{eqnarray*}
	Since $\left[ z_{m-k}', \cdots, z_{m-1}' , z_m'\right]' = M_m v$, the latter equality yields to
	\begin{eqnarray*}
		\liminf\limits_{m \to \infty} \norm{M_m v}{} =0.
	\end{eqnarray*}
	Now, taking an arbitrary $\epsilon>0$, it suffices to show that
	\begin{eqnarray} \label{infinitelyoften}
	\PP{ \inf\limits_{ \norm{v}{}=1} \norm{M_m v}{} < \epsilon , \text{ i.o. for $m$} }=0.
	\end{eqnarray}
	Remember that $L_{m-k}, \cdots, L_m$ are all random matrices thanks to the randomizations $\randommatrix_{m-k}, \cdots, \randommatrix_m$ being used by RCE adaptive regulator. Further, since the distributions of $\randommatrix_{m-k}, \cdots, \randommatrix_m$ are absolutely continuous with respect to Lebesgue measure, we have $\rank{\estA{t}}=p$, for all $t=1,2, \cdots$. So, Lemma \ref{optmanifold} implies that for all $m-k \leq i \leq m$, 
	\begin{eqnarray*}
		\dimension{}{\left\{ \para: \Lmatrix{\para}=L_i \right\}}=p^2.
	\end{eqnarray*}
	Consider the set of matrices $M_m$ such that there exists a vector $v \in \R^q$ to satisfy $\norm{v}{}=1$, as well as $M_m v =0$. For a fixed
	\begin{eqnarray*}
		v=\left[v_1',v_2'\right]', v_1 \in \R^p, v_2 \in \R^r,
	\end{eqnarray*}
	the equality $M_mv=0$ implies $L_i' v_2= -v_1$, for $m-k \leq i \leq m$; i.e. every $L_i$ belongs to a $p(r-1)$ dimensional shifted linear subspace. Putting all above together, the set of $p \times q$ matrices $\para_1, \cdots, \para_{k+1}$ such that there exists some $v$ satisfying $\left[ I_p , \Lmatrix{\para_i}' \right] v=0$ for all $1 \leq i \leq k+1$ is of the dimension 
	\begin{eqnarray*}
		d_1= q-1+(k+1)p^2 +(k+1)p (r-1). 
	\end{eqnarray*}
	Denote the set above by $\mathcal{X} \subset \R^{(k+1)p \times q}$. On the other hand, the set of all $p \times q$ matrices $\para_1, \cdots, \para_{k+1}$ is of the dimension 
	\begin{eqnarray*}
		d_2= (k+1) p q. 
	\end{eqnarray*}
	Now, for $1 \leq i \leq k+1$, suppose that $\para_i$ is the parameter estimate after episode $m-i+1$:
	\begin{eqnarray*}
		\para_i = \estpara{\lfloor \rrate^{m-i+1} \rfloor}.
	\end{eqnarray*}
	So, according to the definition of $M_m$ in \eqref{keymatrixeq}, the inequality $\inf\limits_{ \norm{v}{}=1} \norm{M_m v}{} < \epsilon$, implies that the $(k+1)p \times q$ dimensional matrix
	\begin{eqnarray*}
		\left[m^{1/4}\randommatrix_m, \cdots, \left(m-k\right)^{1/4}\randommatrix_{m-k}\right]
	\end{eqnarray*}
	belongs to an $ \epsilon$-neighborhood of a $d_1 = \dimension{}{\mathcal{X}}$ dimensional set. Since $k$ is sufficiently large to satisfy $d_2-d_1 \geq 5$, we get 
	\begin{eqnarray}
	\PP{ \inf\limits_{ \norm{v}{}=1} \norm{M_m v}{} < \epsilon }= \order{m^{-5/4} \epsilon^5}.
	\end{eqnarray}
	Applying Borel-Cantelli Lemma, we get the desired result in \eqref{infinitelyoften}.
\end{pf*}
\begin{pf*}{Proof of Lemma \ref{stochterm}}
	First, note that 
	\begin{eqnarray*}
		&& \limsup\limits_{m \to \infty} m^{-1} \rrate^{-m/2} S_m \\
		&\leq& \limsup\limits_{m \to \infty} m^{-1/2}\sum\limits_{i=0}^{m} \rrate^{-i/2} \Mnorm{\randommatrix_{m-i}}{}^2, \\
		&\leq& \limsup\limits_{m \to \infty} m^{-1/2} \sum\limits_{i=0}^{m^{1/2}} \Mnorm{\randommatrix_{m-i}}{}^2 \\
		&+& \limsup\limits_{m \to \infty} \rrate^{-m^{1/2}/2} \sum\limits_{i=m^{1/2}}^m  \Mnorm{\randommatrix_{m-i}}{}^2.
	\end{eqnarray*}
	Since $\rrate^{m^{1/2}/2} = \orderinv{m}$, we get
	\begin{eqnarray*}
		&& \limsup\limits_{m \to \infty} m^{-1} \rrate^{-m/2} S_m \\
		&\leq& \limsup\limits_{m \to \infty} m^{-1/2} \sum\limits_{i=0}^{m^{1/2}} \Mnorm{\randommatrix_{m-i}}{}^2 \\
		&+& \limsup\limits_{m \to \infty} m^{-1} \sum\limits_{i=m^{1/2}}^m  \Mnorm{\randommatrix_{m-i}}{}^2.
	\end{eqnarray*}
	Applying the Law of Large Numbers, according to \eqref{perturbmomentcondition} both above terms are $\order{1}$, which is the desired result. A similar discussion holds for $T_m$. 
\end{pf*}

\begin{pf*}{Proof of Lemma \ref{TScov}}
	For the largest eigenvalue, Lemma \ref{smallcov} implies that $\eigmax{\Sigma_m} = \order{\rrate^m}$. To prove of the desired result on the smallest eigenvalue of $\Sigma_m$, we use the approach developed in the proof of Lemma \ref{bigcov}. 
	
	For $i=0,1,\cdots$, let $v_i \in \R^q$ be the eigenvector corresponding to the smallest eigenvalue of $\Sigma_i$. Further, define $\randommatrix_i= \left(\estpara{\lfloor \rrate^i \rfloor} - \mu_i\right) \Sigma_i^{1/2}$. Note that according to the structure of TS, every row of $\randommatrix_i$ is a standard normal (i.e. mean zero Gaussian with covariance $I_q$). We examine the effect of the randomization $\Sigma_i^{-1/2} \randommatrix_i$ on $\Lmatrix{\estpara{\lfloor \rrate^i \rfloor}}$. 
	
	First, we have
	\begin{eqnarray*}
		\Mnorm{\estpara{\lfloor \rrate^i \rfloor} - \mu_i}{} \geq  \randommatrix_i \Sigma_i^{-1/2} v_i = \eigmin{\Sigma_i}^{-1/2} \norm{\randommatrix_i v_i}{}.
	\end{eqnarray*}
	Note that $\randommatrix_i v_i$ is a random vector satisfying \begin{eqnarray*}
		\norm{\randommatrix_i v_i}{} &=& \orderinv{i^{-3/2}}, \\
		\norm{\randommatrix_i v_i}{} &=& \order{i^{1/2}}.
	\end{eqnarray*}
	Then, according to
	\begin{eqnarray*}
		\left(\estpara{\lfloor \rrate^i \rfloor} - \mu_i\right) \Sigma_i \left(\estpara{\lfloor \rrate^i \rfloor} - \mu_i\right)' = \randommatrix_i \randommatrix_i',
	\end{eqnarray*} 
	since $\Mnorm{\randommatrix_i}{}=\order{i^{1/2}}$, Lemma \ref{smallcov} implies that for $j<i$,
	\begin{eqnarray} \label{TScoveq1}
	\Mnorm{\left(\estpara{\lfloor \rrate^i \rfloor} - \mu_i\right) \extendedLmatrix{\estpara{\lfloor \rrate^j \rfloor}}}{} = \order{\rrate^{-j/2}i^{1/2}}.
	\end{eqnarray}
	Letting 
	\begin{eqnarray*}
		D_j &=& \estpara{\lfloor \rrate^j \rfloor} \extendedLmatrix{\estpara{\lfloor \rrate^j \rfloor}}, \\
		Z &=& \Kmatrix{\estpara{\lfloor \rrate^j \rfloor}} \left( \estpara{\lfloor \rrate^i \rfloor} - \mu_i \right) \extendedLmatrix{\estpara{\lfloor \rrate^j \rfloor}}, \\
		\Delta &=& \sum\limits_{t=0}^\infty {D_j'}^t \left( Z' D_j + D_j' Z \right) {D_j}^t,
	\end{eqnarray*}
	\eqref{TScoveq1} implies that $\Mnorm{Z}{}= \order{\rrate^{-j/2}i^{1/2}}, \Mnorm{\Delta}{} = \order{\rrate^{-j/2}i^{1/2}}$. Hence, using \eqref{tangentspace} for $\estpara{\lfloor \rrate^j \rfloor}$, if $j \geq i-k$ for some constant $k$, the following holds:
	\begin{eqnarray} \label{TScoveq2}
		\Mnorm{ \Lmatrix{\estpara{\lfloor \rrate^i \rfloor}} - \Lmatrix{\estpara{\lfloor \rrate^j \rfloor}} }{} = \orderinv{ \frac{\norm{\randommatrix_i v_i}{}}{\eigmin{\Sigma_i}^{1/2}}},
	\end{eqnarray}
	as long as 
	\begin{eqnarray*}
		\limsup\limits_{i \to \infty} \frac{\rrate^{-i/2}i^{1/2}}{\eigmin{\Sigma_i}^{-1/2} \norm{\randommatrix_i v_i}{}} =0.
	\end{eqnarray*}
	
	To proceed, denote the feedback matrix of episode $i$ by $L_i$; i.e. $L_i = \Lmatrix{\estpara{\lfloor \rrate^i \rfloor}}$. Suppose that $k$ is sufficiently large to satisfy $(k+1)p \geq q+3$, and define the $(k+1)p \times q$ matrix 
	\begin{eqnarray*} 
	M_m= \begin{bmatrix}
	(m-k)^{1/4} \rrate^{(m-k)/4} \left[ I_p , L_{m-k}' \right]\\
	\vdots \\
	(m-1)^{1/4} \rrate^{(m-1)/4} \left[ I_p, L_{m-1}' \right] \\
	m^{1/4} \rrate^{m/4} \left[ I_p, L_m' \right] 
	\end{bmatrix}.
	\end{eqnarray*}
	Then, on the event $\eigmin{\Sigma_m} \neq \orderinv{\rrate^{m/2}m^{-1/2}}$, for an arbitrary $\epsilon>0$, the following holds for infinitely many values of $m$:
	\begin{eqnarray} \label{infinitelymanym}
	\inf\limits_{ \norm{v}{}=1} \norm{M_m v}{} < \epsilon . 
	\end{eqnarray}
	Let $\mathcal{Y} \subset \R^{(k+1)p \times r}$ be the set of matrices $\left[L_{m-k}' , \cdots, L_m' \right]$ such that $M_mv=0$, for some unit vector $v \in \R^q$. One can see that
	\begin{eqnarray*}
		d_1 = \dimension{}{\mathcal{Y}}= q-1+ (k+1) p (r-1).
	\end{eqnarray*}
	Whenever \eqref{infinitelymanym} holds, $\left[L_{m-k}' , \cdots, L_m' \right]$ belongs to an $\order{m^{-1/4}\rrate^{-m/4}\epsilon}$-neighborhood of $\mathcal{Y}$. Thus, \eqref{TScoveq2} leads to
	\begin{eqnarray*}
		\PP{\inf\limits_{ \norm{v}{}=1} \norm{M_m v}{} < \epsilon} = \order{\left( m^{-1/2} \epsilon \right)^{(k+1)pr - d_1}}.
	\end{eqnarray*}
	By the choice of $k$, the above terms are summable. So, Borel-Cantelli Lemma implies that with probability one, \eqref{infinitelymanym} can not hold for infinitely many $m$.
\end{pf*}

\end{document}